\documentclass[prd,letterpaper,twocolumn,preprintnumbers,nofootinbib]{revtex4}

\usepackage{slashed}
\usepackage{amsmath,amssymb}
\usepackage{graphicx}
\usepackage{units}
\usepackage{bbold}
\usepackage{xcolor}
\usepackage{dsfont}
\usepackage[hyperfootnotes=false,colorlinks,citecolor=blue]{hyperref}

\usepackage{pifont}

\usepackage{comment}
\usepackage[normalem]{ulem}     

\usepackage{float}

\newcommand{\beq}{\begin{equation}}
\newcommand{\eeq}{\end{equation}}
\newcommand{\bea}{\begin{eqnarray}}
\newcommand{\ena}{\end{eqnarray}}
\newcommand{\dd}{{\rm d}}

\def \epsilon {\varepsilon} 

\def \vec#1{{\boldsymbol{#1}}}

\newcommand{\tr}{\ensuremath{\text{tr}}}

\newcommand{\avb}[1]{\big\langle #1 \big\rangle}	
\newcommand{\eq}{\mathrm{eq}}
\newcommand{\ie}{\emph{i.e.}}

\newcommand{\diff}{\mathrm{d}}
\newcommand{\D}{\mathrm{d}}
\newcommand{\Neff}{N_\mathrm{eff}}

\newcommand{\cmark}{\ding{51}}%
\newcommand{\xmark}{\ding{55}}%

\allowdisplaybreaks

\begin{document}

\title{Testing Dirac leptogenesis with the cosmic microwave background and proton decay}

\author{Julian Heeck}
\email[E-mail: ]{heeck@virginia.edu}
\thanks{ORCID: \href{https://orcid.org/0000-0003-2653-5962}{0000-0003-2653-5962}.}
\affiliation{Department of Physics, University of Virginia,
Charlottesville, Virginia 22904-4714, USA}

\author{Jan Heisig}
\email[E-mail: ]{heisig@virginia.edu}
\affiliation{Department of Physics, University of Virginia,
Charlottesville, Virginia 22904-4714, USA}

\author{Anil Thapa}
\email[E-mail: ]{wtd8kz@virginia.edu}
\thanks{ORCID: \href{https://orcid.org/0000-0003-4471-2336}{0000-0003-4471-2336}.}
\affiliation{Department of Physics, University of Virginia,
Charlottesville, Virginia 22904-4714, USA}

\hypersetup{
pdftitle={Testing Dirac leptogenesis with the cosmic microwave background and proton decay},
pdfauthor={Julian Heeck, Jan Heisig, Anil Thapa}}

\begin{abstract}
The nature of neutrino masses and the matter--antimatter asymmetry of our universe are two of the most important open problems in particle physics today and are notoriously difficult to test with current technology. Dirac neutrinos offer a solution through a leptogenesis mechanism that hinges on the smallness of neutrino masses and resultant non-thermalization of the right-handed neutrino partners in the early universe. We thoroughly explore possible realizations of this Dirac leptogenesis idea, revealing new windows for highly efficient asymmetry generation. In many of them, the number of relativistic degrees of freedom, $N_\text{eff}$, is severely enhanced compared to standard cosmology and offers a novel handle to constrain Dirac leptogenesis with upcoming measurements of the cosmic microwave background. Realizations involving leptoquarks even allow for low-scale post-sphaleron baryogenesis and predict proton decay. These novel aspects render Dirac leptogenesis surprisingly testable.
\end{abstract}

\maketitle

\section{Introduction}
\label{sec:intro}

Neutrino oscillations have established neutrinos to be massive particles, albeit much lighter than all other fermions: $m_\nu \lesssim \unit[0.8]{eV}$~\cite{KATRIN:2021uub}.
The Standard Model (SM) of particle physics needs to be extended by additional particles to accommodate non-zero $m_\nu$, the simplest extension being three right-handed neutrinos $\nu_R$ that form massive Dirac particles together with the familiar $\nu_L$. This is sufficient to explain all neutrino data and makes the mass generation for leptons analogous to that of quarks. The Higgs then couples to neutrinos with coupling strength $m_\nu/\unit[174]{GeV}$, too feeble to be detectable in experiments or even to ever thermalize the $\nu_R$ in the early universe~\cite{Shapiro:1980rr,Antonelli:1981eg,Chen:2015dka}, assuming a vanishing abundance after the Big Bang. Only an undetectably small $\nu_R$ abundance is created through the Higgs interactions~\cite{Adshead:2020ekg,Luo:2020fdt}.

The tiny $\nu_R$ coupling and consequent non-thermalization was used to great effect in Ref.~\cite{Dick:1999je} for \emph{Dirac leptogenesis}: Similar to standard leptogenesis~\cite{Fukugita:1986hr,Davidson:2008bu}, a \emph{lepton} asymmetry is created in the early universe through the  decay of new heavy particles that is then converted to the observed \emph{baryon} asymmetry by sphalerons~\cite{Kuzmin:1985mm}. Where standard leptogenesis creates the lepton asymmetry through explicit lepton number violation, Dirac leptogenesis creates two exactly opposite lepton asymmetries for left- and right-handed neutrinos. Since the latter are invisible to the sphalerons inside the SM plasma, only the left-handed asymmetry is converted into baryons.
The asymmetry within the $\nu_R$, and indeed the $\nu_R$ themselves, are seemingly impossible to observe.

In this article, we provide an exhaustive list of Dirac leptogenesis realizations and study their phenomenology. By solving the relevant Boltzmann equations we show that this mechanism is far more efficient than previously estimated. Furthermore, we show that large regions of parameter space are surprisingly testable or already excluded by measurements of $N_\text{eff}$ in cosmic microwave background (CMB) experiments, going far beyond earlier estimates~\cite{Murayama:2002je,Abazajian:2019oqj}. 
Realizations involving leptoquarks do not even require sphalerons and can thus work at low scales, unavoidably generating proton decay as a consequence.

The rest of this article is structured as follows: in Sec.~\ref{sec:ingredients} we describe the ingredients necessary for Dirac leptogenesis and describe the mechanism qualitatively. In Sec.~\ref{sec:model} we study the simplest realization quantitatively to confirm the qualitative picture from before.
Sec.~\ref{sec:proton_decay} is devoted to a discussion of qualitatively different Dirac-leptogenesis realizations that do not require sphalerons and simultaneously generate proton decay. We conclude in Sec.~\ref{sec:conclusions}.
Some technical details and additional information have been relegated to the appendix: the details of our $N_\text{eff}$ calculation can be found in App.~\ref{sec:Neff_computation} and the full derivation of our Boltzmann equations in App.~\ref{sec:boltzmann}. App.~\ref{sec:cross_sections} lists the relevant scattering cross sections for the model discussed in the main text. 
We illustrate some numerical solutions to the Boltzmann equations in App.~\ref{sec:evolution}.

\section{Ingredients for Dirac Leptogenesis}
\label{sec:ingredients}

\begin{table*}[]
    \centering
    \begin{tabular}{|c|c|c|c|c|c|c|c|c|}
    \hline
     Case & $SU(3)\times SU(2)\times U(1)$ & spin  & $g_X$ & $(B-L)(X)$ & Relevant Lagrangian terms that induce $X$ decay & $\epsilon_\text{wave}$ & $\epsilon_\text{vertex}$ & $\Delta B$ \\ \hline
       $a$ &  $(\vec 1,\vec 1,-1)$ & $0$ & $1$ & $-2$ & $\nu_R e_R \bar{X},\ LL\bar{X} $ & \cmark & \xmark & 0 \\  \hline
       $b$ &  $(\vec 1,\vec 2,1/2)$ & $0$ & $2$ & $0$ & $\bar{H}X,\ \bar{\nu}_R L X,\ \bar{L} e_R X,\ \bar{Q}_L d_R X,\ \bar{u}_R Q_L X,\ X^\dagger H^\dagger H H$ & \cmark & \cmark or \xmark & 0 \\ \hline
      $c$ &   $(\vec 3,\vec 1,-1/3)$ & $0$ & $3$ & $-2/3$ & $d_R \nu_R X^\dagger,\ u_R e_R X^\dagger,\ Q_L L X^\dagger, u_R d_R X,\ Q_L Q_L X$ & \cmark & \cmark or \xmark & \ 0 or 1 \\ \hline
      $d$ &   $(\vec 3,\vec 1,2/3)$ & $0$ & $3$ & $-2/3$ & $u_R \nu_R X^\dagger,\ d_R d_R X$ & \cmark & \xmark & 1 \\ \hline
      $e$ &   $(\vec 3,\vec 2,1/6)$ & $0$ & $6$ & $4/3$ &$\bar{Q}_L \nu_R X,\ \bar{d}_R L X$ & \cmark & \xmark & 0 \\ \hline \hline
      $f$ &   $(\vec 1,\vec 2,-1/2)$ & $1/2$ & $2$ & $-1$ &$\bar{X} L,\ \bar{\nu}_R X H,\ \bar{X} e_R H$ & \cmark & \cmark & 0\\ \hline
    \end{tabular}
    \caption{Quantum numbers for particle $X$ whose decay gives Dirac leptogenesis. $\epsilon_\text{wave}$ and $\epsilon_\text{vertex}$  indicate one-loop contributions from wave-function  and vertex renormalization.  
    Case $c$ and $d$ can lead to $\Delta B = 1$ proton decay (last column).
     }
    \label{tab:quantum_numbers}
\end{table*}

For the simplest Dirac leptogenesis setup, we need several copies of a heavy new particle $X$ that decays -- typically before sphaleron freeze-out, but at least before Big Bang nucleosynthesis -- into a non-thermalized $\nu_R$ plus an SM particle. Since $\nu_R$ is a spin-$1/2$ gauge singlet, $X$ carries the same gauge quantum numbers as the SM particle but has a different spin. Borrowing language  from supersymmetry, $X$ is hence either a  slepton, a  squark, or a Higgsino. Consequently, any supersymmetric Dirac-neutrino model automatically provides the necessary ingredients for Dirac leptogenesis. The different quantum number assignments for $X$ are listed in Tab.~\ref{tab:quantum_numbers}; case $b$ is the one originally discussed in Ref.~\cite{Dick:1999je}.
The same models were identified in Ref.~\cite{Fong:2013gaa} as interesting additions to Majorana-neutrino leptogenesis.
In all cases we can consistently assign a conserved  $B-L$ quantum number to $X$, which allows us to protect the Dirac nature of neutrinos by imposing $U(1)_{B-L}$~\cite{Heeck:2014zfa} or a subgroup~\cite{Heeck:2013rpa} either globally or locally.
$X$ always has additional decay modes exclusively into SM particles  besides the one into $\nu_R$. These are crucial for Dirac leptogenesis because otherwise we would have an additional global $U(1)_{\nu_R}$ symmetry that would lead to CP conservation.

Assuming hierarchical $X$, a CP asymmetry $\epsilon$ in the decays of the lightest $X$ will create a $\nu_R$ asymmetry
\begin{align}
Y_{\Delta \nu_R} 
\equiv \frac{n_{\nu_R}-n_{\bar \nu_R}}{s}\Big|_\text{today} 
\equiv \epsilon \eta \frac{n^\eq_X+n^\eq_{\bar X}}{s}\Big|_{T\gg M_X} 
\end{align}
for every multiplet component of $X$, 
where $s\propto g_\star T^3$ is the entropy density and $n_A$ the number density of $A$.
The above corresponds to the standard definition of the efficiency factor $\eta$~\cite{Hambye:2012fh}, which is however \emph{not} restricted to $|\eta|\leq 1$ for Dirac leptogenesis, although we have $|\epsilon\eta|\leq 1$.

Following the $\nu_R$ asymmetry generation through $X$ decays, the $\nu_R$ will be out of contact with the SM plasma. Since  $B-L$ is conserved in our Dirac-neutrino model, we have the asymmetry $Y_{\Delta (B-L_\text{SM})} =  Y_{\Delta \nu_R}$ in the SM bath. There, sphalerons break $\Delta (B+L_\text{SM}) = 6$, converting the $B-L_\text{SM}$ asymmetry into a baryon asymmetry~\cite{Kuzmin:1985mm,Harvey:1990qw}
\begin{align}
    Y_{\Delta B} = \frac{28}{79}Y_{\Delta (B-L_\text{SM})}= \frac{28}{79}Y_{\Delta \nu_R}\simeq 10^{-3}g_X \epsilon \eta\,,
\end{align}
where in the last equation we have assumed only the SM degrees of freedom, $g_\star = 106.75$, as we will in all numerical examples.
To obtain the measured baryon asymmetry~\cite{Davidson:2008bu,Planck:2018vyg} we thus need $g_X \epsilon \eta\sim 10^{-7}$.
In cases $c$ and $d$, $X$ decays can \emph{directly} produce a baryon asymmetry and can be effective \emph{after} sphaleron freeze out; the factor $28/79$ then needs to be dropped and no $\Delta L_\text{SM}$ is generated.

With CP asymmetry $\epsilon$ simply generated at one loop in all cases, the only quantity left to calculate is the efficiency $\eta$.
From Tab.~\ref{tab:quantum_numbers} it is clear that in addition to the desired decay channels, $X$ unavoidably also has gauge interactions, which can quickly deplete the number of $X$ at temperatures $T< M_X$. Naively, this makes it more complicated to  generate the baryon asymmetry since it suppresses $\eta$. However, in analogy to scalar-triplet leptogenesis~\cite{Hambye:2005tk} there are ways to have very efficient leptogenesis as long as at least some of the inverse decay reactions are out of equilibrium, as already observed in~\cite{Berbig:2022pye}. 

Depending on the hierarchy of rates, different predictions for  $N_\text{eff}$ emerge:
\begin{enumerate}
\item[(I)] If all decay rates of $X$ are out of equilibrium, we have to rely on gauge interactions to produce~$X$, assuming  zero initial abundance. Once these scatterings freeze out, the remaining $X$ eventually decay perfectly out of equilibrium at a temperature $T \ll M_X$. 
The $\nu_R$ created in this decay then have a large momentum compared to the SM temperature and thus a potentially large contribution to $N_\text{eff}$, reminiscent of the superWIMP mechanism~\cite{Decant:2021mhj}, see App.~\ref{sec:Neff_computation}. 
This novel observation severely restricts this region of parameter space.
\item[(II)] If the decay rates involving $\nu_R$ are in equilibrium but the other ones are not, a large $\eta$ can be achieved in complete analogy to scalar-triplet leptogenesis. Here, the $\nu_R$ are thermalized  at $T\sim M_X$, yielding
\begin{align}
\Delta N_\text{eff} \simeq 0.14\, \left(106.75/g_\star (M_X)\right)^{4/3},
\label{eq:thermal_Neff}
\end{align}
an amount testable by CMB-S4~\cite{Abazajian:2019eic} unless $g_\star (M_X)$ far exceeds the SM amount~\cite{Abazajian:2019oqj}.
This is the same contribution as in the $\Delta L = 4$ Dirac-leptogenesis mechanism of Ref.~\cite{Heeck:2013vha}.
\item[(III)] If the decay rates involving $\nu_R$ are out of equilibrium but the other ones are not, we have efficient asymmetry generation with only a small amount of $\nu_R$ generated through freeze-in with typical momenta $p\sim 2.5\, T$~\cite{Heeck:2017xbu}. Here, $\Delta N_\text{eff}$ can be unobservably small since both abundance and momenta of $\nu_R$ are small. This \emph{freeze-in} Dirac leptogenesis technically differs from the namesake setup of~\cite{Li:2020ner}.
\end{enumerate}
The above cases allow for large $\eta$. Moving away from these extreme cases lowers $\eta$ and often pushes $\Delta N_\text{eff}$ closer to the thermal value of Eq.~\eqref{eq:thermal_Neff}.
The interactions and decays of the heavier $X$ copies -- required to exist for non-zero $\epsilon$ -- will further increase $\Delta N_\text{eff}$ without contributing to the asymmetry. Even  case (III) could therefore generate a testable $\Delta N_\text{eff}$ unless the $\nu_R$ couplings of \emph{all} $X$ are suppressed.

Below we quantify the above points  for case $a$ (cf. Tab.~\ref{tab:quantum_numbers}), arguably the simplest version of Dirac leptogenesis. The other cases give qualitatively similar phenomenology, except for the leptoquark cases $c$ and $d$, which are discussed in more detail toward the end.

\section{A simple model}
\label{sec:model}

As a simple model that realizes Dirac leptogenesis we introduce two electrically charged scalars $X_{1,2}\equiv X^-_{1,2}$ to the SM (case $a$ from Tab.~\ref{tab:quantum_numbers}), in addition to the three right-handed neutrinos necessary to form Dirac neutrinos. The Yukawa couplings of $\nu_R$ with the Higgs are minuscule and play no role in the following. 
The relevant interactions of the charged scalars are
\begin{equation}
    {\cal L} =  \tfrac{1}{2}\bar{L}^c F_{i} L\ \bar{X}_i + \bar{e}^c G_i \nu_{R} \bar{X}_i + {\rm  h.c.} \, ,
\end{equation}
assuming, without loss of generality, that the $X_i$ are mass eigenstates.
The matrices $F_{1,2}$ are antisymmetric in their flavor indices due to the antisymmetry of the $SU(2)$ singlet contraction $ \bar{L}^c_\alpha L_\beta = \bar{e}^c_\alpha \nu_{L,\beta}-\bar{e}^c_\beta \nu_{L,\alpha}$.
The $G_i$ are arbitrary complex Yukawa matrices. Total lepton number is conserved by assigning $L(X_i)=2$, but, more importantly, $\nu_R$ number is explicitly broken by the simultaneous presence of both Yukawas; this allows for the generation of a $\nu_R$ asymmetry in the $X_i$ decays.\footnote{These couplings contribute at one-loop level to the Dirac-neutrino mass matrix, $\delta m_\nu \sim F m_\ell G^* v^2/(16 \pi^2 M_X^2) $, which is $\ll m_\nu$ in the region of interest without finetuning.}

\begin{figure}
    \centering
    \includegraphics[scale=0.40]{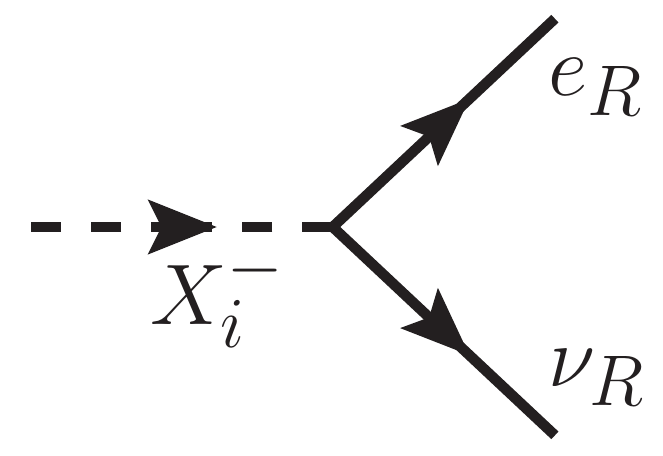}\hspace{10mm}
    \includegraphics[scale=0.40]{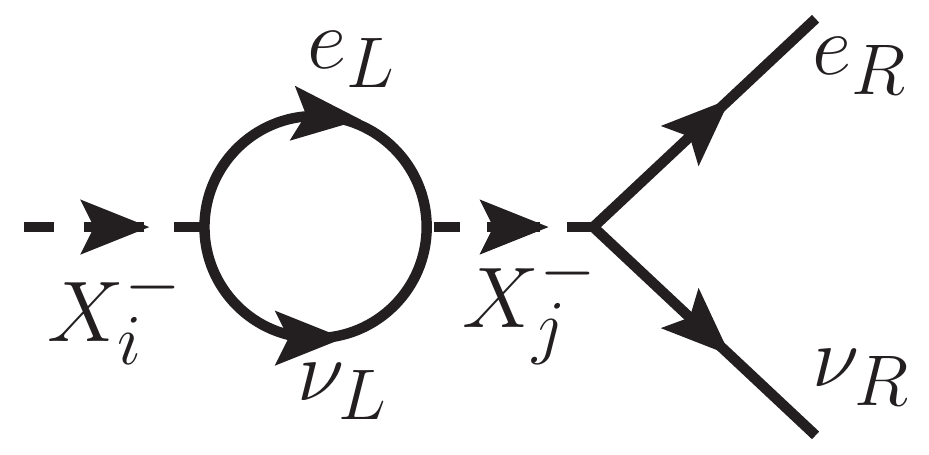}
    \caption{Tree-level and one-loop wave-function diagram whose interference produces the CP-asymmetry $\epsilon_i$ of Eq.~\eqref{eq:one-loop_epsilon}.
    }
    \label{fig:self-energy}
\end{figure}

The tree-level decay rates of $X_i$ are given by
\begin{align}
    \Gamma (X_i\to e_{R} \nu_{R}) &\simeq \frac{M_{i}}{16\pi} \tr(G_iG_i^\dagger)\,,\\
    \Gamma (X_i\to e_{L} \nu_{L}) &\simeq \frac{M_{i}}{16\pi} \tr(F_iF_i^\dagger)\,,
\end{align}
summed over all final-state flavors. CPT invariance enforces $ \Gamma(X_i) = \Gamma (\bar{X}_i)\equiv\Gamma_{X_i}$ and hence 
\begin{align}
    \Gamma (X_i\to e_{R/L} \nu_{R/L}) & = \Gamma_{X_i}\left( B_{R/L} \pm \epsilon_i\right), \\
    \Gamma (\bar{X}_i\to\bar e_{R/L} \bar \nu_{R/L}) & = \Gamma_{X_i}\left( B_{R/L} \mp \epsilon_i\right)
\end{align}
in the presence of a CP asymmetry 
\begin{align}
    \epsilon_i  &\equiv \frac{\Gamma (X_i \to e_
R \nu_R) -\Gamma (\bar X_i \to \bar{e}_R \bar{\nu}_R)}{2 \Gamma_{X_i}} \,,
\end{align}
where $B_L = 1-B_R$ and $B_R\equiv \Gamma (X_i\to e_{R} \nu_{R})/\Gamma_{X_i}$.
This definition of $\epsilon_i$ as the average $\nu_R$ number per $X_i$ decay immediately implies the absolute upper bound
\begin{align}
|\epsilon_i |\leq \text{min}(B_L,B_R)\,,
\label{eq:largest_epsilon}
\end{align}
although realistic values for $\epsilon_i$ are far below this limit.
This is in complete analogy to triplet-scalar leptogenesis~\cite{Hambye:2005tk}.
At one loop, we find from the diagrams in Fig.~\ref{fig:self-energy}:
\begin{align}
    \epsilon_i 
    &\simeq \frac{\sum_{j}  \left(M_j^2/M_i^2-1\right)^{-1} \Im\ [\tr(F_i F_j^\dagger) \tr(G_i G_j^\dagger)]}{8 \pi\ [\tr(G_i G_i^\dagger)+\tr(F_i F_i^\dagger)]} \,.
    \label{eq:one-loop_epsilon}
\end{align}
Again: a non-vanishing CP asymmetry unavoidably requires both decay modes $X_i\to e_L\nu_L$ and  $X_i\to e_R\nu_R$.

In the following we will assume a hierarchical $X_i$ spectrum  with $X_1\equiv X$ being the lightest. Any asymmetry generated by the heavier $X_{i>1}$ is expected to be washed out by the interactions of $X$; the contribution of the heavier scalars to the $\nu_R$ density and thus $\Delta N_\text{eff}$ on the other hand will only increase the final $\Delta N_\text{eff}$. By neglecting these contributions here we are being conservative.

Provided all leptons except for $\nu_R$ are in thermal equilibrium, the Boltzmann equations (derived in App.~\ref{sec:boltzmann}) for $\Sigma_A\equiv Y_A+Y_{\bar A}$ and $\Delta_A\equiv Y_A-Y_{\bar A}$ read
\begin{align}
  &\frac{\dd \Sigma_X}{\dd y} = 
  \frac12 \avb{\sigma v}_{X\bar X}\left(\Sigma_X^2- {\Sigma_X^{\eq}}^2\right)\label{eq:BME:SigmaX}\\
&\qquad\qquad + \frac{\langle \Gamma_X\rangle}{s} \left[\Sigma_X-\Sigma_X^{\eq} \left(B_L+B_R \frac{\Sigma_{\nu_R}}{\Sigma_{\nu_R}^\eq}\right)\right],\nonumber
\\
\label{eq:BME:SigmanuR}
 &\frac{\dd \Sigma_{\nu_R}}{\dd y} =
  - \frac{\langle \Gamma_X\rangle}{s}  B_R \left(\Sigma_X - \Sigma_X^\eq\frac{\Sigma_{\nu_R}}{\Sigma_{\nu_R}^\eq}\right)\\
  & \qquad\qquad  +\frac12 \left(\avb{\sigma v}_s \Sigma_{e_R}^\eq
 +\avb{\sigma v}_t \Sigma_{L}^\eq\right)\left( \Sigma_{\nu_R}-\Sigma_{\nu_R}^\eq\right)  ,\nonumber
  \\
 &\frac{\dd \Delta_X}{\dd y} =
 \frac{\langle \Gamma_X\rangle}{s}  \bigg[
 \Delta_X - \Sigma_X^\eq \bigg\{
 B_R \frac{\Delta_{\nu_R}\left( \Sigma_{e_R}^\eq + \Sigma_{\nu_R} \right)}{\Sigma_{\nu_R}^\eq \Sigma_{e_R}^\eq}\label{eq:BME:DeltaX}\\
 &\qquad \qquad\quad \;\;\,
 - B_L \frac{4 \left(\Delta_X + \Delta_{\nu_R}\right)}{\Sigma_L^\eq}
 -\epsilon \left( 1- \frac{\Sigma_{\nu_R}}{\Sigma_{\nu_R}^\eq} \right)
  \bigg\}
   \bigg],\nonumber
 \\
 &\frac{\dd \Delta_{\nu_R}}{\dd y} =
 \frac{\langle \Gamma_X\rangle}{s}  \bigg[
- \epsilon \left( \Sigma_X - \Sigma_X^\eq \frac{\Sigma_{\nu_R}}{\Sigma_{\nu_R}^\eq}\right)
\label{eq:BME:DeltanuR}\\
&\qquad \qquad \qquad\;  - B_R\bigg(  \Delta_X - \Sigma_X^\eq \frac{\Delta_{\nu_R}\left( \Sigma_{e_R}^\eq + \Sigma_{\nu_R} \right)}{\Sigma_{\nu_R}^\eq \Sigma_{e_R}^\eq} \bigg)
  \bigg]\nonumber  \\
& 
\quad\;+\frac{\avb{\sigma v}_s}{2} \left[
\Delta_{\nu_R} \left( \Sigma_{e_R}^\eq + \Sigma_{\nu_R}\right)
+ 2 \left( \Delta_{\nu_R} +\Delta_{X} \right) \Sigma_{\nu_R}^\eq 
\right] \nonumber \\
&
\quad\;\;\;+\avb{\sigma v}_t \left[
\Delta_{\nu_R}  \Sigma_{{ L}}^\eq
+  \left(  \Delta_{\nu_R} +\Delta_{X}  \right)\left(\Sigma_{\nu_R}+ \Sigma_{\nu_R}^\eq\right)
\right] ,\nonumber
\end{align}
 where $\dd/\dd y \equiv 3 \mathcal{H}\, (\dd s/\dd x)^{-1} \dd/\dd x$, 
$\mathcal{H}$ is the Hubble rate and $x \equiv M_X/T$.
Neglecting the Yukawa interactions as well as couplings in the scalar potential, the relevant thermally averaged $X$--$\bar X$ annihilation cross section $\avb{\sigma v}_{X\bar X}$ comes from the hypercharge coupling of $X$  see e.g.~Ref.~\cite{Cirelli:2007xd}.
The $s$- and $t$-channel $X$-mediated $|\Delta \nu_R|=1$ scatterings such as $e_L\nu_L\to e_R\nu_R$ are encoded in $\avb{\sigma v}_{s,t}$ and are typically suppressed compared to the (inverse) decays. 
Details can be found in App.~\ref{sec:cross_sections}.

\begin{figure}
    \centering
    \includegraphics[width=0.48\textwidth]{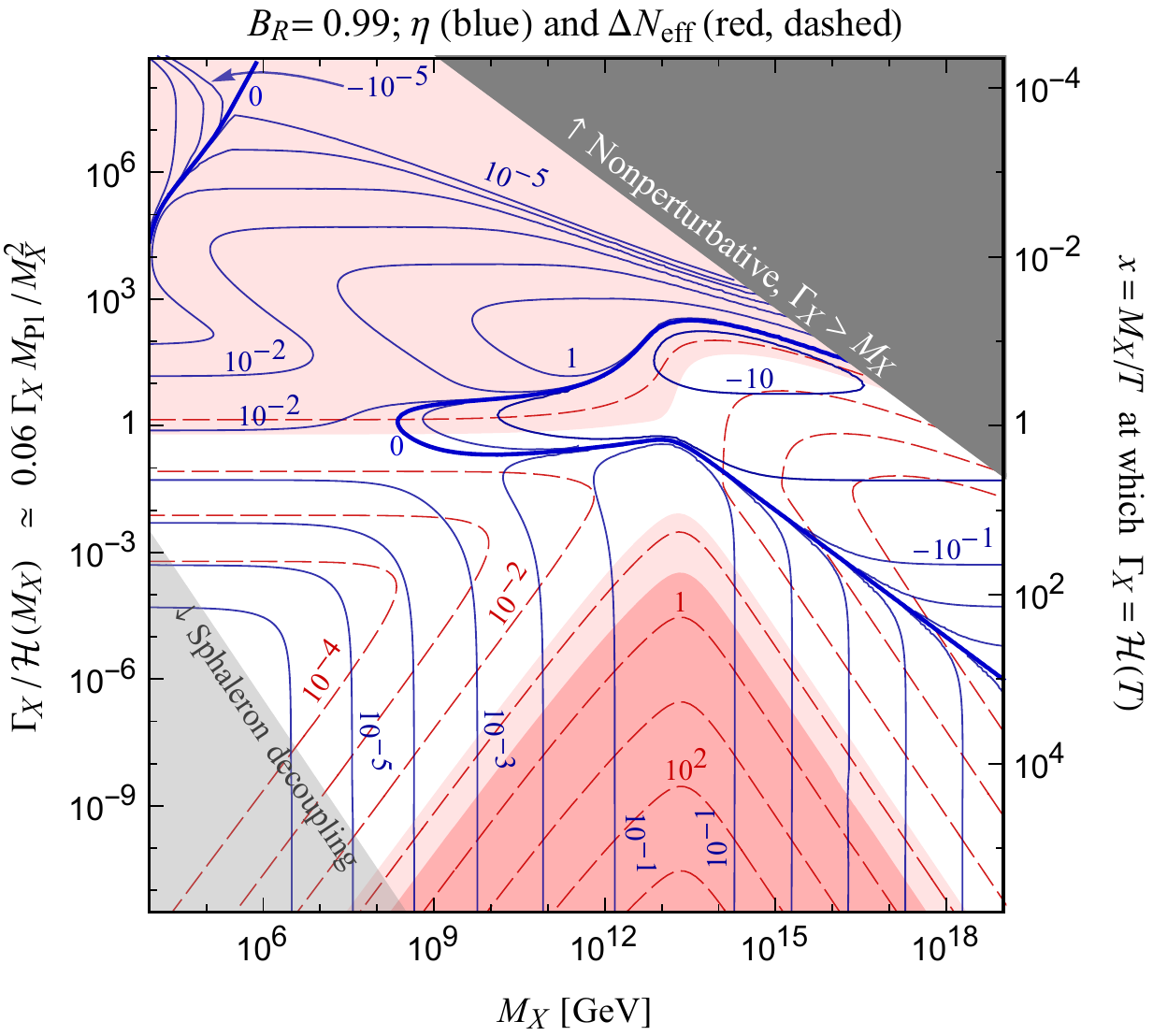}\\
    \vspace{1.3ex}
    \includegraphics[width=0.48\textwidth]{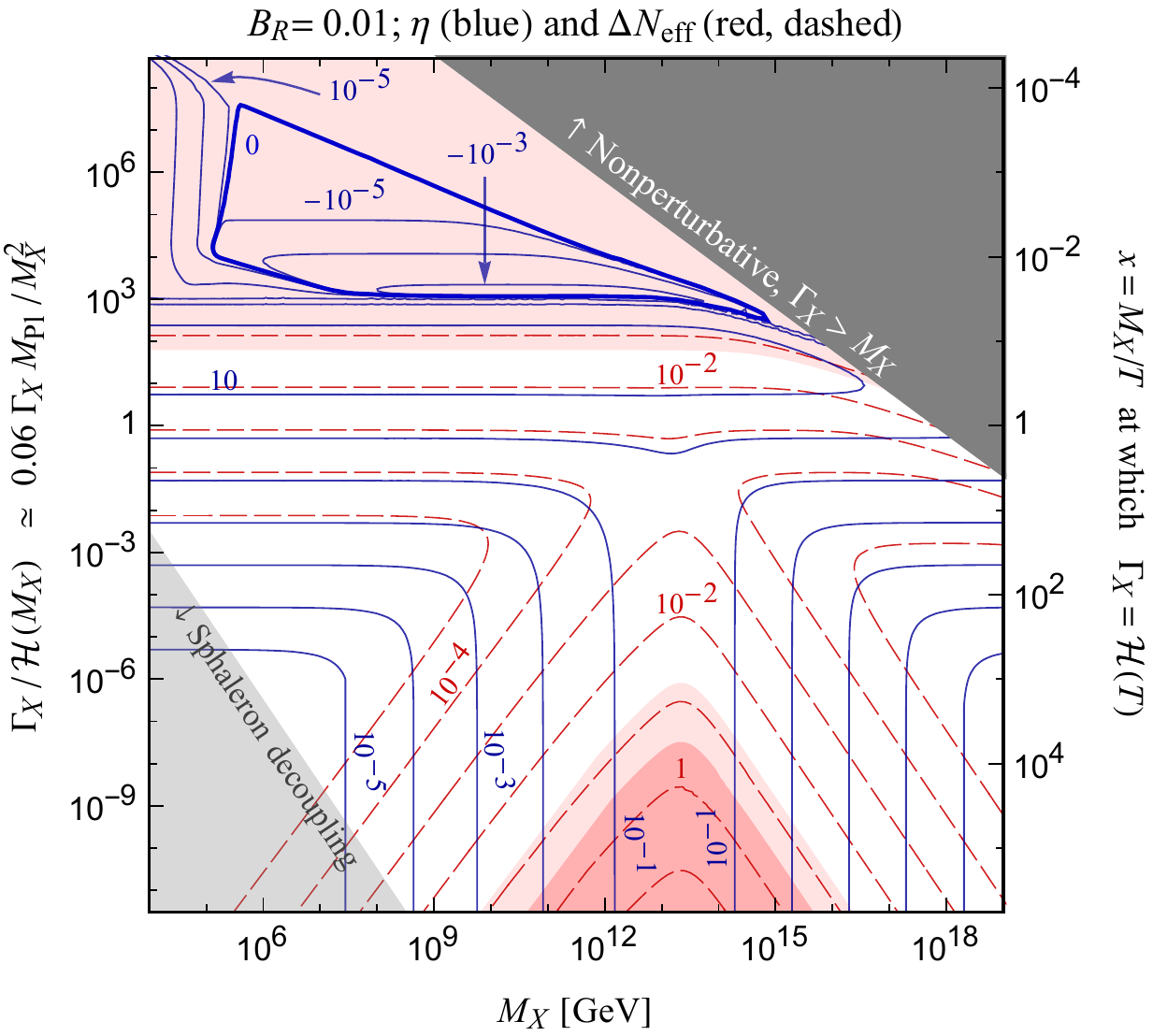}
    \caption{Contours of $\Delta N_\text{eff}$ (red, dashed) and $\eta$ (blue) for $B_R=0.99$ (top) and $B_R=0.01$ (bottom). In both cases $|\epsilon| \leq 10^{-2}$ and we require $\epsilon \eta \sim 10^{-7}$ to explain the baryon asymmetry.
The dark red shaded region is excluded by current bounds on $\Delta \Neff$~\cite{Planck:2018vyg}; light red shows future reach~\cite{Abazajian:2019eic}.
The upper-right region has nonperturbative $\Gamma_X > M_X$. In the lower-left region, $X$ decays after sphaleron decoupling.
    }
    \label{fig:contours}
\end{figure}

The above set of Boltzmann equations has already been simplified by setting the linear combinations $\Delta_{\nu_R} - \Delta_{e_R} $ and $\Delta_{\nu_R} + \Delta_{e_R}+\Delta_{\nu_L} + \Delta_{e_L} +2\Delta_{X}$, which are conserved due to $U(1)_Y\times U(1)_L$, to zero, and by assuming all $\Delta$ to be suppressed by the small $\epsilon$. We assume vanishing initial abundances for both $X$ and $\nu_R$. For thermalized $\nu_R$, these equations are similar to those of triplet leptogenesis~\cite{Hambye:2005tk}.
We show some numerical solutions to the Boltzmann equations in App.~\ref{sec:evolution} to illustrate the evolution of $\nu_R$ abundance and asymmetry.
$\Delta_{\nu_R} (T\to 0)$ gives the $\nu_R$ asymmetry or efficiency parameter $\eta$, while  $\Sigma_{\nu_R} (T\to 0)$ is the number of right-handed neutrinos, which gives $\Delta N_\text{eff}$ when multiplied by the characteristic $\nu_R$ momentum at production; see appendix~\ref{sec:Neff_computation} for more details.
Some numerical solutions are presented in Fig.~\ref{fig:contours}.

In Fig.~\ref{fig:contours}, we can recognize the behavior mentioned before and can quantify the relations:
\begin{enumerate}
\item[(I)] For $\Gamma_X \ll \mathcal{H}$, the $X$ freeze in or  out and decay at $T \ll M_X$. The efficiency peaks at $M_X\sim \unit[10^{13}]{GeV}$ and then falls off	like $M_X^{-1}$ ($M_X\log (M_X/{\rm GeV})$) for larger (smaller) masses. Here,
\begin{align}
    \Delta N_\text{eff}\simeq 0.05\,B_R \eta /\sqrt{\Gamma_X M_\text{Pl}/M_X^2}
    \label{eq:superWIMP_Neff}
\end{align}
can become arbitrarily large for small $\Gamma_X$ due to the large $\nu_R$ momentum, leading to strong constraints.
A freeze-in component $ \Delta N_\text{eff}\propto \Gamma_X/M_X^2$ becomes important for larger $\Gamma_X$ and eventually leads to thermalization.
\item[(II)] For $B_L\ll 1$ and thermalized $X\to \nu_R$, a large $\eta \lesssim 1/B_L$ can be obtained, although $|\epsilon \eta|$ remains below 1. This is an efficient leptogenesis region with the simple $\Delta N_\text{eff}$ prediction of Eq.~\eqref{eq:thermal_Neff}.
\item[(III)] For $B_R\ll 1$, we find $\Delta_{\nu_R} \simeq \Sigma_{\nu_R} \epsilon/B_R$ and thus a large  $\eta \lesssim 1/B_R$, together with a suppressed $\Delta N_\text{eff}$.
\end{enumerate}

\begin{figure}[!ht]
    \includegraphics[width=0.44\textwidth]{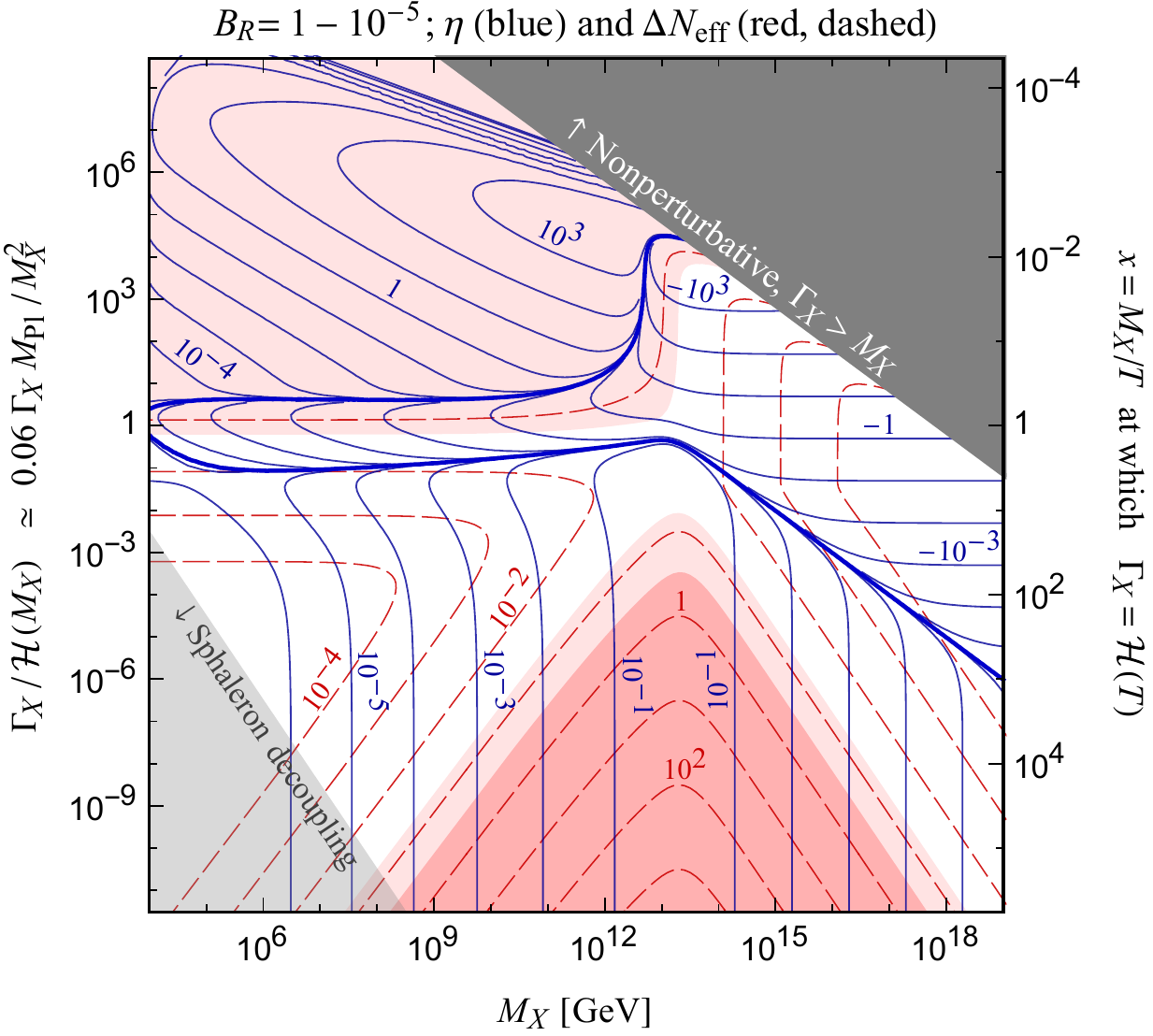}\\[4pt]
    \includegraphics[width=0.44\textwidth]{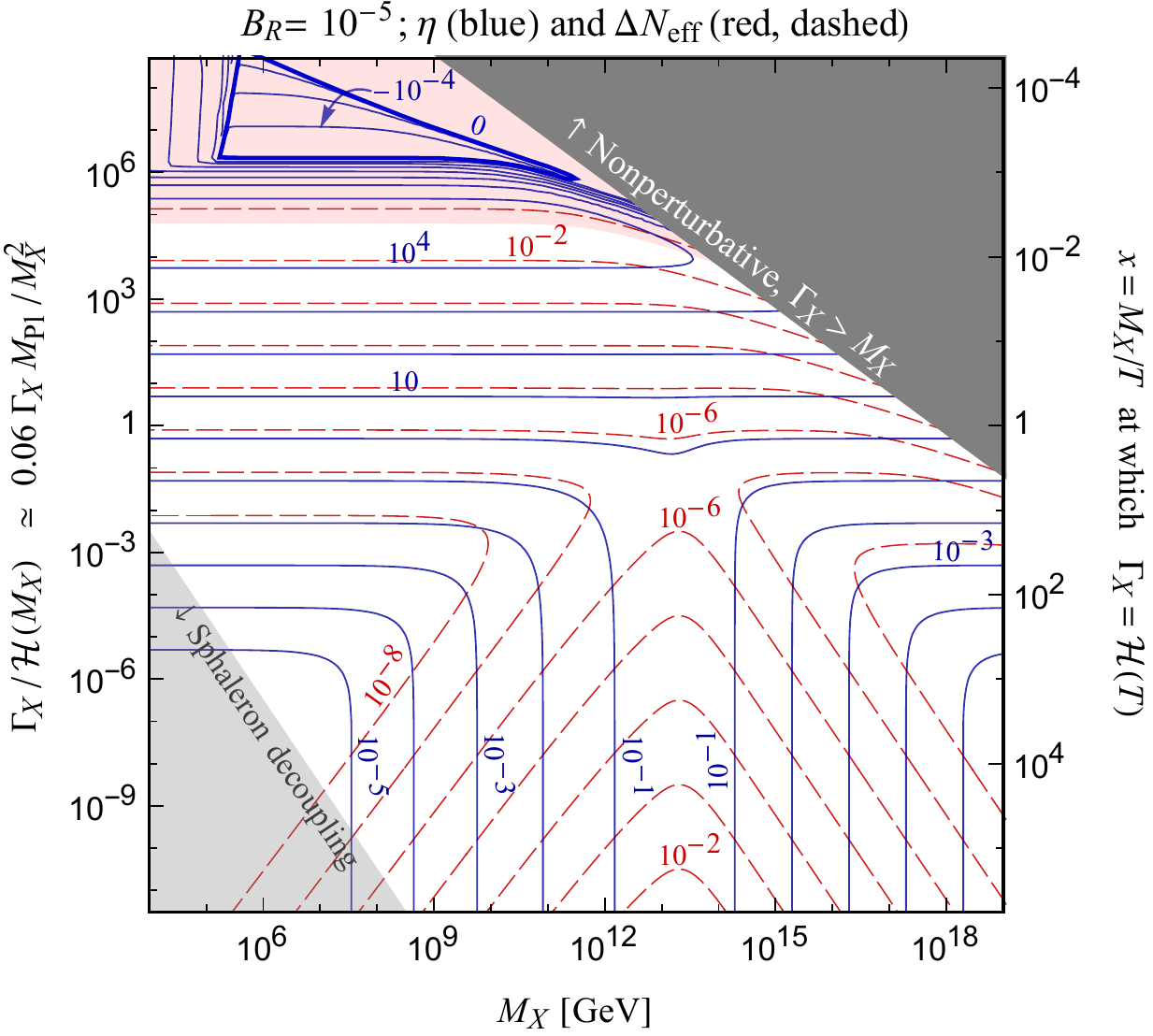}\\[4pt]
    \includegraphics[width=0.44\textwidth]{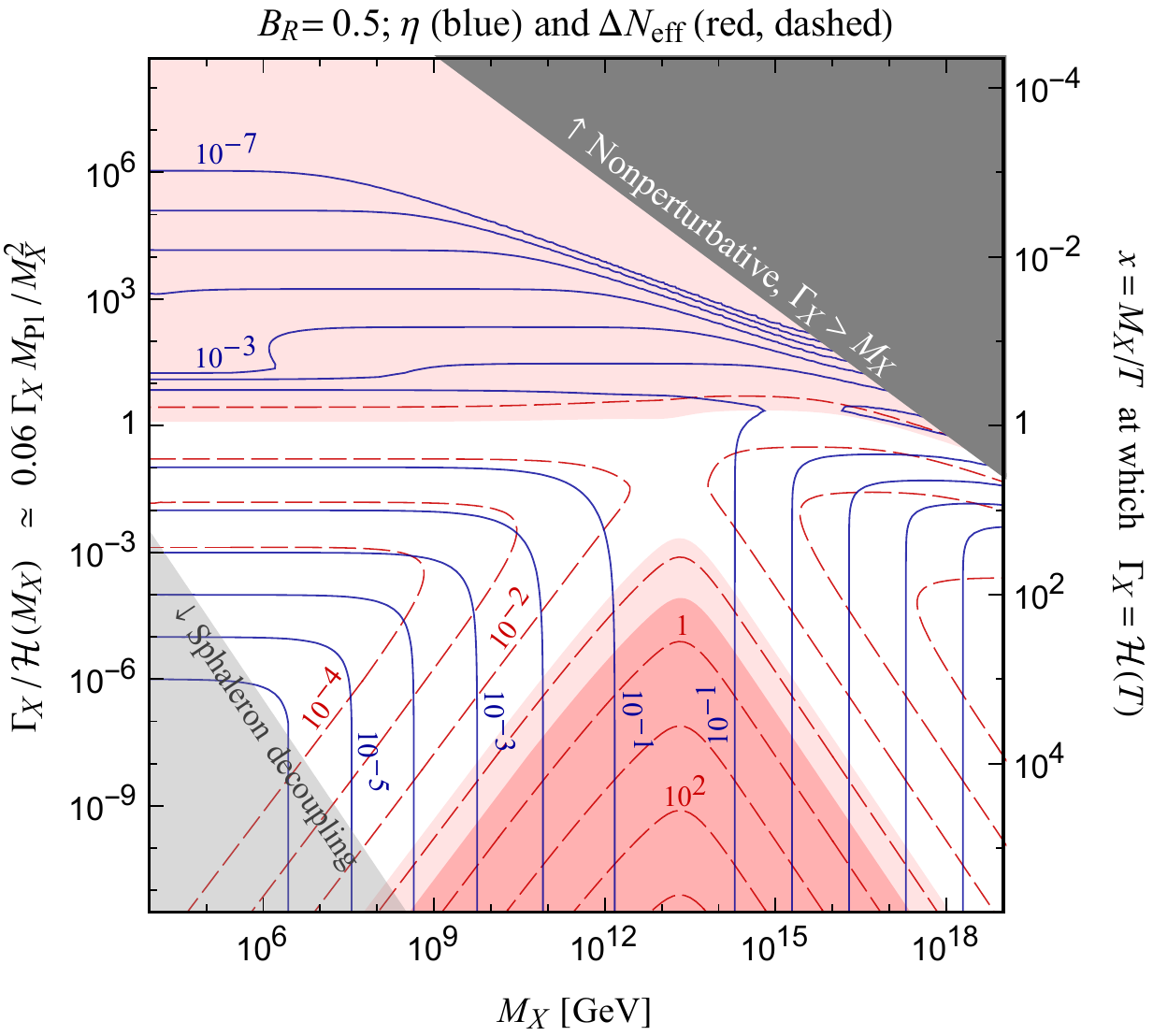}
    \caption{Same as Fig.~\ref{fig:contours} but for $B_R = 1-10^{-5}$ (top), $B_R = 10^{-5}$ (middle), $B_R =1/2$ (bottom).
    }
    \label{fig:contours2}
\end{figure}

Numerical examples for other branching ratios are given in Fig.~\ref{fig:contours2} and confirm the above picture.
The qualitative behavior for the choices $B_R=1-10^{-5}$ (upper panel), and $B_R=10^{-5}$ (middle panel) is similar to that of Fig.~\ref{fig:contours} and we can identify the large-$\eta$ regions already described above.
Notice that in these two examples $|\epsilon|$ is restricted to be below $10^{-5}$ from Eq.~\eqref{eq:largest_epsilon}, so $|\eta|$ has to be larger than $10^{-2}$ in order to generate the observed baryon asymmetry. For $\Gamma_X \ll \mathcal{H}(M_X)$, this restricts $M_X$ to the region $10^{11}$--$\unit[10^{15}]{GeV}$, but the mass is essentially unconstrained in regions (II) and (III), i.e.~for larger $\Gamma_X$.
Realistically, $\epsilon$ is actually much smaller than this upper limit of $10^{-5}$  and thus $\eta$ has to be larger still. Nevertheless, we can have successful baryogenesis over a wide region of parameter space.
In Fig.~\ref{fig:contours2} (bottom), we show the efficiency for the special case $B_R = B_L = 1/2$. This case is comparably simple because the lack of hierarchy in $X\leftrightarrow e_R \nu_R$ vs.~$X\leftrightarrow e_L \nu_L$ precludes the large-$\eta$ regions (II) and (III). $\eta$ can at most be of order one here, which however is hardly restrictive because $|\epsilon|$ can be as large as $1/2$ in principle. Even this case allows therefore for efficient baryogenesis, with a large portion of parameter space testable through $\Delta N_\text{eff}$.

From Figs.~\ref{fig:contours} and~\ref{fig:contours2} it is clear that Dirac leptogenesis is very efficient, in part because both $X$ \emph{and} $\nu_R$ can be out of equilibrium, allowing for successful baryogenesis even with tiny $\epsilon$. Since the sign of $\eta$ depends on the hierarchy of rates we find $\eta=0$ contours that delineate these regions, not found in other leptogenesis mechanisms~\cite{Hambye:2012fh}.
Large regions of the parameter space are already excluded by $\Delta N_\text{eff}$ constraints and even more can be tested with stage-IV CMB data~\cite{Abazajian:2019oqj,Abazajian:2019eic}, down to $\Delta N_\text{eff} \simeq 0.06$.

While we have focused our numerical study on case~$a$, the other cases of Tab.~\ref{tab:quantum_numbers} are qualitatively similar. Their gauge annihilation cross sections will differ somewhat -- and might even require corrections due to  Sommerfeld~\cite{ElHedri:2016onc} and bound-state formation~\cite{Gross:2018zha} -- and there are often more than two relevant decay channels, but the basic picture from, say, Fig.~\ref{fig:contours} remains correct. Let us briefly mention two cases that induce new effects.

\section{Proton decay}
\label{sec:proton_decay}

Case $d$ (and in general case $c$) of Tab.~\ref{tab:quantum_numbers} is special in that it violates baryon number directly. This makes it possible to circumvent the use of sphalerons in baryogenesis and establish \emph{low-scale} Dirac leptogenesis, sharing similarities with cloistered baryogenesis~\cite{AristizabalSierra:2013lyx}. The parameter space looks similar to Fig.~\ref{fig:contours}, except that the lower-left sphaleron decay region is now allowed, $X$ only needs to decay before Big Bang nucleosynthesis. This enlarges the allowed parameter space and in particular allows for fairly light leptoquarks $X$, which could then lead to detectable particle-physics signatures. Interestingly, the fact that a non-zero CP asymmetry $\epsilon$ requires $X$ couplings to both $u_R\nu_R$ and $d_R d_R$ unequivocally  gives rise to proton decay! $B-L$ is conserved in these proton decays and we unavoidably have final states that contain $\nu_R$ (see also~\cite{Helo:2018bgb}). For case $d$, we \emph{only} have such $\nu_R$ final states, e.g.~$p\to K^+ \bar{\nu}_R$, while case $c$ also has fully visible final states such as $p\to \pi^0 e^+$. 

For case $d$, we have the following Lagrangian for several copies of the leptoquark $X\sim (\vec{3},\vec{1},2/3)$,
\begin{align}
    {\cal L} =  \bar{d}^c_R F_{i} d_R\, X_i + \bar{u}^c_R  G_i \nu_{R}\,\bar{X}_i + {\rm  h.c.} \, ,
\end{align}
with implicit contraction of $SU(3)$ indices.
This leads to the proton decay rate
\begin{align}
\Gamma (p\to K^+\bar\nu_R) 
&\simeq \sum_{i,\alpha}\frac{ |F_{i,12}G_{i,1\alpha}|^2}{\unit[6\times 10^{33}]{yr}} \left(\frac{\unit[2\times 10^{15}]{GeV}}{M_{X_i}}\right)^4 ,
\end{align}
using the relevant QCD matrix element from Ref.~\cite{Yoo:2021gql}. Notice that a kaon is produced due to the antisymmetry of the $F_i$ Yukawa couplings in flavor space. The current limit on this proton-decay mode is $1/\Gamma >\unit[6\times 10^{33}]{yr}$~\cite{Super-Kamiokande:2014otb} and will be improved in JUNO~\cite{JUNO:2022qgr}, Hyper-Kamiokande~\cite{Hyper-Kamiokande:2022smq} and DUNE~\cite{DUNE:2020fgq}.
Baryogenesis does not actually require $X$ couplings to the first quark generation, seemingly allowing for an easy way to evade proton decay. However, \emph{any} nonzero $F$ and $G$ couplings will together -- as required for a nonzero CP asymmetry -- induce proton decay at higher order in perturbation theory, potentially with more complicated final states~\cite{Heeck:2019kgr}. The $X$ masses and couplings required for baryogenesis can easily lead to testable proton decay rates (and $\Delta N_\text{eff}$).

\section{Conclusions}
\label{sec:conclusions}

Massive Dirac neutrinos have Higgs couplings too small to bring the $\nu_R$ into thermal equilibrium, which allows for leptogenesis without $B-L$ violation.
In this article, we have shown that there are many simple realizations of this two-decade-old idea and that each one has a much larger viable parameter space than anticipated: Dirac leptogenesis is very efficient. Even more surprisingly, much of this parameter space is testable through the $\nu_R$ contribution to $N_\text{eff}$, soon to be measured with sub-percent accuracy by CMB stage-IV experiments. A subset of models even allows for post-sphaleron baryogenesis and predict proton decay, making them one of the few known models that link these two baryon number violating observables.
With both baryogenesis and Dirac neutrinos notoriously difficult to probe, Dirac leptogenesis provides some novel handles for testability.

\section*{Acknowledgements}

This work was supported in part by the National Science Foundation under Grant PHY-2210428.
J.~Heisig acknowledges support by the Alexander von Humboldt foundation via the Feodor Lynen Research Fellowship for Experienced Researchers. 
We acknowledge Research Computing at the University of Virginia for providing computational resources that have contributed to the results reported within this publication.

\appendix

\section{Computation of \texorpdfstring{$\Delta N_\text{eff}$}{Delta Neff}}
\label{sec:Neff_computation}

At temperature $T$, the energy density of the universe can be written as
\begin{equation}
\rho = \left[ 1 + \frac78 \left( \frac{T_\nu}{T}\right)^4 \left(\Neff + \Delta \Neff\right) \right]  \rho_\gamma  + \dots ,
\end{equation}
where $\rho_\gamma$ is the energy density of photons, $\Neff= 3.045$~\cite{deSalas:2016ztq} is the SM's effective number of relativistic degrees of  freedom in the active neutrino sector and
\begin{equation}
\Delta \Neff =\frac87\left(\frac{T}{T_\nu}\right)^4 \frac{\rho_i}{\rho_\gamma}  
\end{equation}
is the respective contribution from an additional relativistic species $i$ with energy density $\rho_i$. 
In general,
\begin{align}
\rho_i = g_i \int \frac{\diff^3 p_i}{(2 \pi)^3} \, E_i f_i\,,
\end{align}
where $g_i$ is the particle's number of internal degrees of freedom, $E_i$ its energy and $f_i$ its momentum distribution.
For ultra-relativistic particles, $E_i=p_i$ and we can express the energy density as $\rho_i = s^{4/3} \langle q_i\rangle Y_i$, where 
$\langle q_i\rangle$ is the first moment of the momentum mode, $q_i\equiv p_i/s^{1/3}$,
\begin{equation}
\langle q_i\rangle = 	\frac{g_i }{s^{4/3} Y_i} \int \frac{\diff^3 p_i}{(2 \pi)^3} \, p_i f_i\,,
\end{equation}
and $Y_i$ is the comoving number density,
\begin{equation}
Y_i= \frac{g_i}{s}  \int \frac{\diff^3 p_i}{(2 \pi)^3} \,  f_i\,.
\end{equation}
In the above expressions, $s = g_* T^3 \,2\pi^2/45$ denotes the entropy density.
Accordingly, 
\begin{equation}
\label{eq:NeffqY}
\Delta \Neff =  \sum_i \frac{\langle  q_i \rangle}{\langle  q_\nu \rangle } \frac{Y_i}{Y_\nu} \Neff\,,
\end{equation}
where $\langle q_\nu\rangle$ and $Y_\nu$ are the respective quantities for the relativistic SM neutrinos:
\begin{align}
\langle q_\nu\rangle &\simeq 3.15\left(\frac{45}{2 \pi^2 g_* (T_{\nu,\text{FO}})}\right)^{1/3} ,
\label{eq:thermalq}
\\
 Y_\nu &= \sum_{j=1}^3 Y_{\nu_j}^\eq(T_{\nu,\text{FO}}) \,,
\end{align}
with $T_{\nu,\text{FO}}$ being the temperature of neutrino decoupling.

In Eq.~\eqref{eq:NeffqY}, the sum runs over the involved production modes of right-handed neutrinos with characteristic $\langle q_i \rangle$ for which we employ the results of Ref.~\cite{Decant:2021mhj}.
For the production in the late decay of the mother particle $X$ (referred to as the superWIMP production mechanism in~\cite{Decant:2021mhj}) we obtain
\begin{equation}
\langle q_\text{SW}\rangle = \frac{45^{1/12} \pi^{7/12}}{2^{4/3} g_*^{1/12}}\frac{M_X}{ \sqrt{\Gamma_{X} M_\text{Pl}}}
\end{equation}
in our notation.
Early production around $x\sim 1$ (via freeze-in or close-to-equilibrium processes) gives rise to a moment similar to Eq.~\eqref{eq:thermalq}.
The respective contributions to the comoving number density, $Y_i$, are obtained from solving the Boltzmann equations.

\section{Boltzmann equations}
\label{sec:boltzmann}

In this appendix we derive the Boltzmann equations. For the individual abundances $Y$ of particles and antiparticles, the 
Boltzmann equations read:
\begin{widetext}
\begin{align}
    \label{eqBMEYX}
  &\frac{\D Y_X}{\D x} = \frac{1}{ 3 \mathcal{H}}\frac{\D s}{\D x}
  \Bigg[
  \avb{\sigma v}_{X\bar X}\left(Y_X Y_{\bar X}- {Y_X^{\eq}}^2\right)
+ \frac{\langle \Gamma_X\rangle \left(B_R +\epsilon \right)}{s}\left(Y_X-Y_X^{\eq}\frac{Y_{\nu_R} Y_{e_R} }{Y_{\nu_R}^\eq Y_{e_R}^\eq}\right)
\nonumber \\ 
&\hspace{17ex}
+\frac{\langle \Gamma_X\rangle \left(B_L - \epsilon\right)}{s}\left(Y_X-Y_X^{\eq}\frac{{Y_{L} }^2 }{{Y_{L}^\eq}^2}\right)
\Bigg],
\\
\label{eqBMEYXbar}
  &\frac{\D Y_{\bar X}}{\D x} = \frac{1}{ 3 \mathcal{H}}\frac{\D s}{\D x}
  \Bigg[
  \avb{\sigma v}_{X\bar X}\left(Y_X Y_{\bar X}- {Y_X^{\eq}}^2\right)
+ \frac{\langle \Gamma_X\rangle \left(B_R -\epsilon \right)}{s}\left(Y_{\bar X}-Y_X^{\eq}\frac{Y_{\bar \nu_R} Y_{\bar  e_R} }{Y_{\nu_R}^\eq Y_{e_R}^\eq}\right)
\nonumber \\ 
&\hspace{17ex}
+\frac{\langle \Gamma_X\rangle \left(B_L + \epsilon\right)}{s}\left(Y_{\bar X}-Y_X^{\eq}\frac{{Y_{\bar L} }^2 }{{Y_{L}^\eq}^2}\right)
\Bigg],
\\
\label{eqBMEYnuR}
  &\frac{\D Y_{\nu_R}}{\D x} = \frac{1}{ 3 \mathcal{H}}\frac{\D s}{\D x}
  \Bigg[
- \, \frac{\langle \Gamma_X\rangle \left(B_R +\epsilon \right)}{s}\left(Y_X-Y_X^{\eq}\frac{Y_{\nu_R} Y_{e_R} }{Y_{\nu_R}^\eq Y_{e_R}^\eq}\right)
+\avb{\sigma v}_{\nu_R e_R \to L L }\left(Y_{\nu_R} Y_{e_R} - Y_{\nu_R}^\eq Y_{e_R}^\eq\frac{ Y_L^2}{{Y_L^\eq}^2} \right)\nonumber \\ 
&\hspace{17ex}+\,\avb{\sigma v}_{\nu_R \bar L \to \bar e_R L }\left(Y_{\nu_R} Y_{\bar L} - Y_{\nu_R}^\eq Y_{L}^\eq\frac{ Y_{\bar e_R} Y_L}{Y_{e_R}^\eq Y_L^\eq} \right)
\Bigg],
\\
\label{eqBMEYnuRbar}
  &\frac{\D Y_{\bar \nu_R}}{\D x} = \frac{1}{ 3 \mathcal{H}}\frac{\D s}{\D x}
  \Bigg[
- \frac{\langle \Gamma_X\rangle \left(B_R -\epsilon \right)}{s}\left(Y_{\bar X}-Y_X^{\eq}\frac{Y_{\bar \nu_R} Y_{\bar  e_R} }{Y_{\nu_R}^\eq Y_{e_R}^\eq}\right)
+\avb{\sigma v}_{\bar\nu_R \bar e_R \to \bar L\bar L }\left(Y_{\bar \nu_R} Y_{\bar e_R} - Y_{\nu_R}^\eq Y_{e_R}^\eq\frac{ Y_{\bar L}	^2}{{Y_L^\eq}^2} \right)\nonumber \\ 
&\hspace{17ex}
+\,\avb{\sigma v}_{\bar\nu_R  L \to e_R\bar  L }\left(Y_{\bar \nu_R} Y_{L} - Y_{\nu_R}^\eq Y_{L}^\eq\frac{ Y_{e_R} Y_{\bar L}}{Y_{e_R}^\eq Y_L^\eq} \right)
\Bigg],
\\
\label{eqBMEYeR}
  &\frac{\D Y_{e_R}}{\D x} = \frac{1}{ 3 \mathcal{H}}\frac{\D s}{\D x}
  \Bigg[
-  \frac{\langle \Gamma_X\rangle \left(B_R +\epsilon \right)}{s}\left(Y_X-Y_X^{\eq}\frac{Y_{\nu_R} Y_{e_R} }{Y_{\nu_R}^\eq Y_{e_R}^\eq}\right)
+\avb{\sigma v}_{\nu_R e_R \to L L }\left(Y_{\nu_R} Y_{e_R} - Y_{\nu_R}^\eq Y_{e_R}^\eq\frac{ Y_L^2}{{Y_L^\eq}^2} \right)
\nonumber \\ 
&\hspace{17ex}
-\avb{\sigma v}_{\bar\nu_R  L \to  e_R \bar L }\left(Y_{\bar \nu_R} Y_{L} - Y_{\nu_R}^\eq Y_{L}^\eq\frac{ Y_{e_R} Y_{\bar L}}{Y_{e_R}^\eq Y_L^\eq} \right) 
+ \mathrm{SM~gauge~interactions}
\;
\Bigg],
\\
\label{eqBMEYeRbar}
  &\frac{\D Y_{\bar e_R}}{\D x} = \frac{1}{ 3 \mathcal{H}}\frac{\D s}{\D x}
  \Bigg[
- \frac{\langle \Gamma_X\rangle \left(B_R -\epsilon \right)}{s}\left(Y_{\bar X}-Y_X^{\eq}\frac{Y_{\bar \nu_R} Y_{\bar  e_R} }{Y_{\nu_R}^\eq Y_{e_R}^\eq}\right)
+\avb{\sigma v}_{\bar\nu_R \bar e_R \to \bar L \bar L }\left(Y_{\bar \nu_R} Y_{\bar e_R} - Y_{\nu_R}^\eq Y_{e_R}^\eq\frac{ Y_{\bar L}	^2}{{Y_L^\eq}^2} \right)
\nonumber \\ 
&\hspace{17 ex}
- \avb{\sigma v}_{\nu_R \bar L \to \bar e_R L }\left(Y_{\nu_R} Y_{\bar L} - Y_{\nu_R}^\eq Y_{L}^\eq\frac{ Y_{\bar e_R} Y_L}{Y_{e_R}^\eq Y_L^\eq} \right)
+ \mathrm{SM~gauge~interactions}
\;
\Bigg],
\\
\label{eqBMEYL}
  &\frac{\D Y_{L}}{\D x} = \frac{1}{ 3 \mathcal{H}}\frac{\D s}{\D x}
  \Bigg[
- 2 \frac{\langle \Gamma_X\rangle \left(B_L - \epsilon\right)}{s}\left(Y_X-Y_X^{\eq}\frac{{Y_{L} }^2 }{{Y_{L}^\eq}^2}\right)
-2\avb{\sigma v}_{\nu_R e_R \to L L }\left(Y_{\nu_R} Y_{e_R} - Y_{\nu_R}^\eq Y_{e_R}^\eq\frac{ Y_L^2}{{Y_L^\eq}^2} \right)
\nonumber \\ 
&\hspace{10ex}
-\avb{\sigma v}_{\nu_R \bar L \to \bar e_R L }\left(Y_{\nu_R} Y_{\bar L} - Y_{\nu_R}^\eq Y_{L}^\eq\frac{ Y_{\bar e_R} Y_L}{Y_{e_R}^\eq Y_L^\eq} \right) 
+\avb{\sigma v}_{\bar\nu_R  L \to  e_R  \bar L }\left(Y_{\bar \nu_R} Y_{L} - Y_{\nu_R}^\eq Y_{L}^\eq\frac{ Y_{e_R} Y_{\bar L}}{Y_{e_R}^\eq Y_L^\eq} \right) + \mathrm{gauge~int.}
\Bigg],
\\
\label{eqBMEYLbar}
  &\frac{\D Y_{\bar L}}{\D x} = \frac{1}{ 3 \mathcal{H}}\frac{\D s}{\D x}
  \Bigg[
- 2 \frac{\langle \Gamma_X\rangle \left(B_L + \epsilon\right)}{s}\left(Y_{\bar X}-Y_X^{\eq}\frac{{Y_{\bar L} }^2 }{{Y_{L}^\eq}^2}\right)
-2 \avb{\sigma v}_{\bar\nu_R \bar e_R \to \bar L \bar L }\left(Y_{\bar \nu_R} Y_{\bar e_R} - Y_{\nu_R}^\eq Y_{e_R}^\eq\frac{ Y_{\bar L}	^2}{{Y_L^\eq}^2} \right)
\nonumber \\ 
&\hspace{10ex}
-\avb{\sigma v}_{\bar\nu_R  L \to  e_R \bar L }\left(Y_{\bar \nu_R} Y_{L} - Y_{\nu_R}^\eq Y_{L}^\eq\frac{ Y_{e_R} Y_{\bar L}}{Y_{e_R}^\eq Y_L^\eq} \right)
+\avb{\sigma v}_{\nu_R \bar L \to \bar e_R L }\left(Y_{\nu_R} Y_{\bar L} - Y_{\nu_R}^\eq Y_{L}^\eq\frac{ Y_{\bar e_R} Y_L}{Y_{e_R}^\eq Y_L^\eq} \right) 
+ \mathrm{gauge~int.}
\Bigg].
\end{align}
\end{widetext}
Notice that $(3 \mathcal{H})^{-1} \D s/\D x = -s/(\mathcal{H} x)$ for constant relativistic degrees of freedom.
As we will assume the SM gauge interactions to be fully efficient, we have combined $\nu_L$ and $e_L$ in the above equations by defining $Y_L=Y_{\nu_L}+Y_{e_L}$. 
Now, we define $ \Sigma_A \equiv Y_A + Y_{\bar A}$ and $ \Delta_A \equiv Y_A - Y_{\bar A}$
for any species $A$ and rewrite the Boltzmann equations accordingly. $\Sigma_{L}$ and $\Sigma_{e_R}$ are approximated by their equilibrium values on account of their efficient SM gauge interactions, leaving the following six equations:
\begin{widetext}
\begin{align}
  &\frac{\D \Sigma_X}{\D x} = \frac{1}{ 3 \mathcal{H}}\frac{\D s}{\D x}
  \left[
  \frac12 \avb{\sigma v}_{X\bar X}\left(\Sigma_X^2- {\Sigma_X^{\eq}}^2\right)
+ \frac{\langle \Gamma_X\rangle}{s} \left\{\Sigma_X-\Sigma_X^{\eq} \left(B_L+B_R \frac{\Sigma_{\nu_R}}{\Sigma_{\nu_R}^\eq}\right)\right\}
\right],
\\
 &\frac{\D \Sigma_{\nu_R}}{\D x} =\frac{1}{ 3 \mathcal{H}}\frac{\D s}{\D x} 
 \frac{\langle \Gamma_X\rangle}{s} 
  \left[
  -  B_R \left(\Sigma_X - \Sigma_X^\eq\frac{\Sigma_{\nu_R}}{\Sigma_{\nu_R}^\eq}\right)
  +\frac12 \avb{\sigma v}_s \Sigma_{e_R}^\eq\left( \Sigma_{\nu_R}-\Sigma_{\nu_R}^\eq\right)
 +\frac12 \avb{\sigma v}_t \Sigma_{L}^\eq\left( \Sigma_{\nu_R}-\Sigma_{\nu_R}^\eq\right)
  \right],
  \\
 &\frac{\D \Delta_X}{\D x} =\frac{1}{ 3 \mathcal{H}}\frac{\D s}{\D x}
 \frac{\langle \Gamma_X\rangle}{s}  \left[
 \Delta_X - \Sigma_X^\eq \left\{
 B_R \frac{\Sigma_{\nu_R} \Delta_{e_R} + \Sigma_{e_R}^\eq \Delta_{\nu_R} }{\Sigma_{\nu_R}^\eq \Sigma_{e_R}^\eq}
 + B_L \frac{2 \Delta_L}{\Sigma_L^\eq}
 -\epsilon \left( 1- \frac{\Sigma_{\nu_R}}{\Sigma_{\nu_R}^\eq} \right)
  \right\} 
   \right],
 \\
 &\frac{\D \Delta_{\nu_R}}{\D x} =\frac{1}{ 3 \mathcal{H}}\frac{\D s}{\D x}
 \frac{\langle \Gamma_X\rangle}{s}  \bigg[
- B_R\left(  \Delta_X - \Sigma_X^\eq \frac{\Sigma_{\nu_R} \Delta_{e_R} + \Sigma_{e_R}^\eq \Delta_{\nu_R} }{\Sigma_{\nu_R}^\eq \Sigma_{e_R}^\eq} \right)
- \epsilon \left( \Sigma_X - \Sigma_X^\eq \frac{\Sigma_{\nu_R}}{\Sigma_{\nu_R}^\eq}\right)
\\ 
&\qquad\qquad +
\frac12 \avb{\sigma v}_s \bigg(
\Delta_{\nu_R}  \Sigma_{e_R}^\eq + \Delta_{e_R}  \Sigma_{\nu_R}
-2  \Delta_{L}  \Sigma_{\nu_R}^\eq \frac{\Sigma_{e_R}^\eq}{\Sigma_{L}^\eq}
\bigg)
+\frac12 \avb{\sigma v}_t \bigg\{
\Delta_{\nu_R}  \Sigma_L^\eq 
-  \Delta_{L} \left(\Sigma_{\nu_R}+ \Sigma_{\nu_R}^\eq\right)
+ \Delta_{e_R} \Sigma_{\nu_R}^\eq\frac{\Sigma_{L}^\eq}{\Sigma_{e_R}^\eq}
\bigg\}
  \bigg],\nonumber
 \\
  \label{eq:BME:DeltaeR}
 &\frac{\D \Delta_{e_R}}{\D x} =\frac{1}{ 3 \mathcal{H}}\frac{\D s}{\D x}
 \frac{\langle \Gamma_X\rangle}{s}  \bigg[
- B_R\left(  \Delta_X - \Sigma_X^\eq \frac{\Sigma_{\nu_R} \Delta_{e_R} + \Sigma_{e_R}^\eq \Delta_{\nu_R} }{\Sigma_{\nu_R}^\eq \Sigma_{e_R}^\eq} \right)
- \epsilon \left( \Sigma_X - \Sigma_X^\eq \frac{\Sigma_{\nu_R}}{\Sigma_{\nu_R}^\eq}\right)
  \\ 
&\qquad\qquad +
\frac12 \avb{\sigma v}_s \bigg(
\Delta_{\nu_R}  \Sigma_{e_R}^\eq + \Delta_{e_R}  \Sigma_{\nu_R}
-2  \Delta_{L}  \Sigma_{\nu_R}^\eq \frac{\Sigma_{e_R}^\eq}{\Sigma_{L}^\eq}
\bigg)
+\frac12 \avb{\sigma v}_t \bigg\{
\Delta_{\nu_R}  \Sigma_L^\eq 
-  \Delta_{L} \left(\Sigma_{\nu_R}+ \Sigma_{\nu_R}^\eq\right)
+ \Delta_{e_R} \Sigma_{\nu_R}^\eq\frac{\Sigma_{L}^\eq}{\Sigma_{e_R}^\eq}
\bigg\}
  \bigg],\nonumber
 \\
   \label{eq:BME:DeltaL}
 &\frac{\D \Delta_L}{\D x} =\frac{1}{ 3 \mathcal{H}}\frac{\D s}{\D x}
 \frac{\langle \Gamma_X\rangle}{s}  \bigg[
-2 B_L \left( \Delta_X - 2 \Sigma_X^\eq \frac{\Delta_L}{\Sigma_L^\eq}   \right)
+ 2 \epsilon \left( \Sigma_X - \Sigma_X^\eq \right)
- \avb{\sigma v}_s \bigg(
\Delta_{\nu_R}  \Sigma_{e_R}^\eq + \Delta_{e_R}  \Sigma_{\nu_R}
-2  \Delta_{L}  \Sigma_{\nu_R}^\eq \frac{\Sigma_{e_R}^\eq}{\Sigma_{L}^\eq}
\bigg) \nonumber
\\ 
&\qquad\qquad 
-\avb{\sigma v}_t \bigg\{
\Delta_{\nu_R}  \Sigma_L^\eq 
-  \Delta_{L} \left(\Sigma_{\nu_R}+ \Sigma_{\nu_R}^\eq\right)
+ \Delta_{e_R} \Sigma_{\nu_R}^\eq\frac{\Sigma_{L}^\eq}{\Sigma_{e_R}^\eq}
\bigg\}
  \bigg],
\end{align}
\end{widetext}
where we have only kept terms linear in $\epsilon$, $\Delta_A$ and neglected any asymmetry in the $s$- and $t$-channel scattering cross sections, $\avb{\sigma v}_{\nu_R e_R \to L L } \simeq \avb{\sigma v}_{\bar\nu_R \bar e_R \to \bar L\bar L }$ and $\avb{\sigma v}_{\nu_R \bar L \to \bar e_R L } \simeq \avb{\sigma v}_{\bar\nu_R  L \to e_R\bar  L }$, denoted by $\avb{\sigma v}_s$ and $\avb{\sigma v}_t$, respectively.

Note that $\D (\Delta_{\nu_R}+\Delta_{e_R}+\Delta_L+2 \Delta_X)/\D x = 0 $ and $\D (\Delta_{\nu_R}-\Delta_{e_R})/\D x = 0 $ due to conservation of hypercharge and lepton number, \ie~the set of differential equations is redundant and we can eliminate two of them by plugging in the solutions
\begin{align}
&\Delta_{\nu_R}+\Delta_{e_R}+\Delta_L+2 \Delta_X = \delta_1\,,\\
&\Delta_{\nu_R}-\Delta_{e_R} = \delta_2\,,
\end{align}
where $\delta_i$ are initial conditions (set to zero here assuming vanishing asymmetries in the beginning). We choose to eliminate Eqs.~\eqref{eq:BME:DeltaeR} and \eqref{eq:BME:DeltaL}  to obtain the Boltzmann equations in the main text.
When $\nu_R$ is deep in equilibrium the Boltzmann equations simplify and are structurally similar to those of triplet leptogenesis~\cite{Hambye:2005tk}. In that region the equations become symmetric under $L\leftrightarrow R$, $\epsilon\leftrightarrow -\epsilon$; for every $\eta$ at $(B_R,\epsilon)$ there is a solution with $-\eta$ at $(1-B_R,-\epsilon)$. This can be observed in the upper-left corners of the two examples in Fig.~\ref{fig:contours}.

\section{Involved cross sections}
\label{sec:cross_sections}

\begin{figure}
    \includegraphics[width=0.42\textwidth]{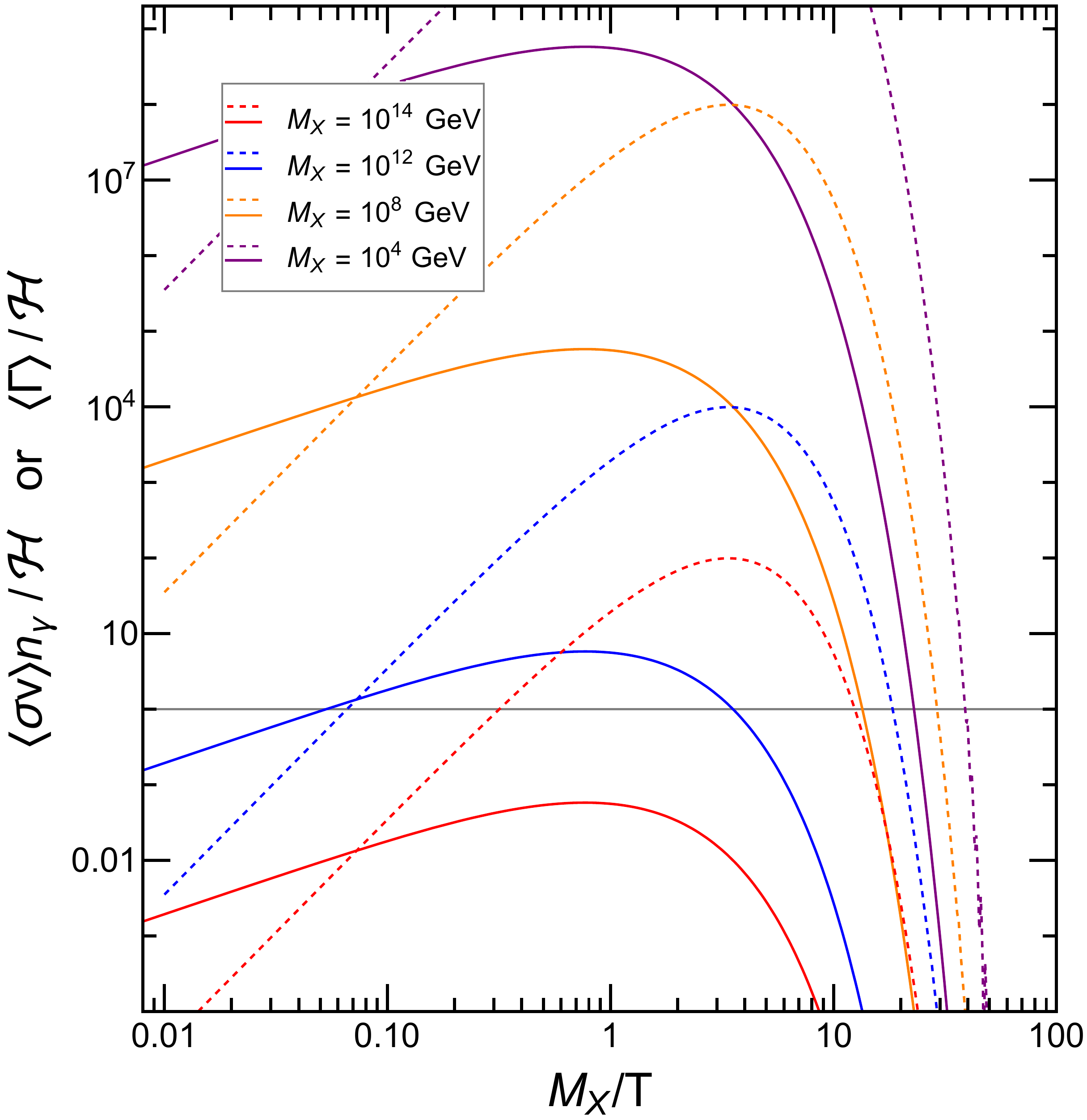}
    \caption{Thermally averaged annihilation rate $\langle \sigma v \rangle n_\gamma/{\cal H}$ (solid lines) and inverse decay rate $\langle \Gamma \rangle/ {\cal H}$ (dashed lines; for unity Yukawas) compared to the Hubble rate for several $X$ masses. 
    }
    \label{fig:rates}
\end{figure}

For case $a$ of Tab.~\ref{tab:quantum_numbers}, the relevant annihilation cross sections, summed over final-state spins, are
\begin{align}
&\sigma (X\bar{X}\to \mathcal{B}^*\to f\bar{f}) = \frac{(g')^4 Y_f^2}{24\pi s}\sqrt{1-\frac{4 M_X^2}{s}}\,,\\
&\sigma (X\bar{X}\to \mathcal{B}^*\to \phi\bar{\phi}) = \frac{(g')^4Y_\phi^2}{48\pi s}\sqrt{1-\frac{4 M_X^2}{s}}\,,\\
&\sigma (X\bar{X}\to \mathcal{B}\mathcal{B}) = \frac{(g')^4}{16\pi M_X^2}\,\Bigg[\frac{\sqrt{(y-1) y^3}+\sqrt{(y-1) y}}{(y-1) y^2}\nonumber\\
&\qquad\qquad\quad 
+ \frac{(1-2 y) \tanh ^{-1}\!\left(\sqrt{\frac{y-1}{y}}\right)}{(y-1) y^2}\Bigg]_{y=\frac{s}{4M_X^2}} ,
\end{align}
where $\mathcal{B}$ is the hypercharge gauge boson and $g'$ the gauge coupling. $f$ is a massless chiral fermion with hypercharge $Y_f$ and $\phi$ a massless complex scalar with hypercharge $Y_\phi$.
The thermally-averaged annihilation rate is approximately
\begin{align}
    \langle \sigma v \rangle_{X\bar X} \simeq \frac{10^{-4}}{M_X^2} \begin{cases}
3.5 \,(M_X/T)^2 \,,& M_X\ll T\,,\\
1.2 \,,& M_X\sim T\,,\\
5.5 \,,& M_X\gg T\,.
    \end{cases}
\end{align}
The thermally-averaged annihilation rate as well as the decay rates are shown in Fig.~\ref{fig:rates} relative to the Hubble rate $\mathcal{H} = \sqrt{4\pi^3 g_*/45} \,T^2/M_\text{Pl}$.
For $M_X\gg \unit[10^{13}]{GeV}$, the hypercharge gauge interactions are not sufficient to put $X$ in equilibrium; for $M_X\ll \unit[10^{13}]{GeV}$, $X$ reaches equilibrium and freezes out at some temperature $T< M_X$. Decay rates have a different temperature dependence than annihilations.

The $X$-mediated $\Delta \nu_R = 1$ scattering cross sections  consist of $s$- and $t$-channel cross sections,
\begin{align}
\sigma_s (L_\alpha L_\beta \to e_{R,\gamma}\nu_{R,\sigma}) &= \frac{|F_{\alpha \beta} G_{\sigma \gamma}^*|^2}{16\pi} \frac{s}{(s-M_X^2)^2}\,,\\
\sigma_t (\bar{L}_\alpha e_{R,\gamma}\to L_{\beta}\bar{\nu}_{R,\sigma}) &= \frac{|F_{\alpha \beta} G_{\sigma \gamma}^*|^2}{16\pi M_X^2}
\\
&\hspace{-11ex}\times\left[\frac{y (y+2)+2 (y+1) \log \left(\frac{1}{y+1}\right)}{y^2 (y+1)}\right]_{y=s/M_X^2}\!\! .\nonumber
\end{align}
The former needs to be properly regulated to subtract the on-shell region that is already counted in the Boltzmann equations via (inverse) decays. We follow the procedure from Ref.~\cite{Hambye:2005tk}
(see also Refs.~\cite{Cline:1993bd,Giudice:2003jh})
and subtract $\sigma_s$ by 
\begin{align}
\frac{|F_{\alpha \beta} G_{\sigma \gamma}^*|^2}{16\pi} \frac{\pi s}{M_X\Gamma_X}\delta (s-M_X^2)\,.
\end{align}
For the Boltzmann equations we require the thermally averaged cross sections, summed over initial and final flavors.
The relevant coupling trace can then also be written as
\begin{align}
\tr(F_iF_i^\dagger)\tr(G_iG_i^\dagger) \simeq B_R B_L \Gamma_X^2 \left(\frac{16\pi}{M_X}\right)^2 .
\end{align}

\section{Evolution of abundances}
\label{sec:evolution}

In this appendix we show some numerical solutions to the Boltzmann equations.
Due to our approximations in App.~\ref{sec:boltzmann}, all $\Delta_A$ are proportional to $\epsilon$ so we are effectively solving for $\Delta/\epsilon$. Depending on the parameters, some $\Delta $  change sign during the evolution.

\begin{figure*}
    \centering
    \includegraphics[scale=0.7]{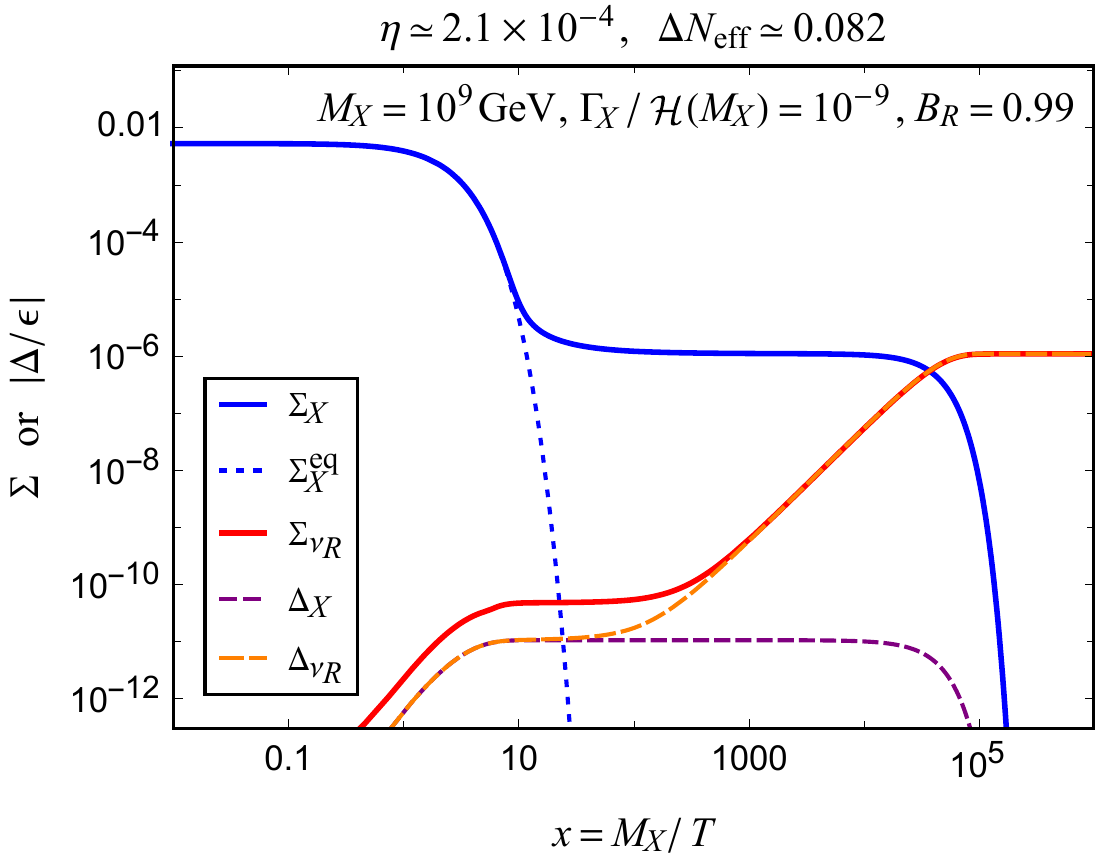} \hspace{10mm}
    \includegraphics[scale=0.7]{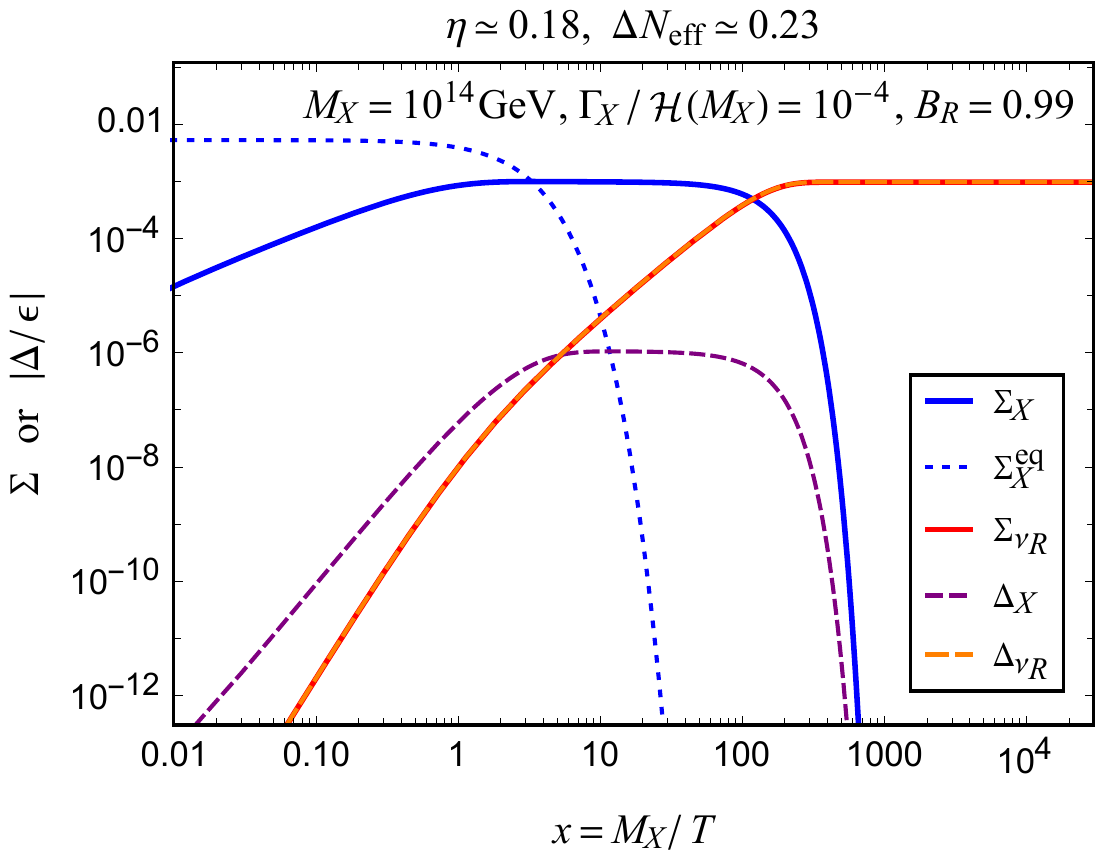}\\
    \caption{Evolution of $\Sigma_{X,\nu_R}$ and $|\Delta_{X,\nu_R}/\epsilon|$ for two parameter points together with the resulting $\eta$ and $\Delta N_\text{eff}$. The dashed blue line shows the equilibrium distribution of $\Sigma_X$. In the right panel, the curves for $\Sigma_{\nu_R}$ and $|\Delta_{\nu_R}/\epsilon|$ are on top of each other.
    }
    \label{fig:evolution1}
\end{figure*}

The two plots in Fig.~\ref{fig:evolution1} correspond to case (I), where $X$ reaches equilibrium and freezes out (left) or freezes in (right) due to its gauge interactions, then decays at $T\ll M_X$. 
The smaller mass in the left figure results in a (more) efficient annihilation, cf.~Fig.~\ref{fig:rates}, leaving few $X$ to eventually decay, which results in a suppressed $\eta$. 
In the right figure, the gauge interactions of $X$ are just too small to thermalize $X$ but still large enough to copiously produce $X$ available to decay, resulting in a large $\eta$. For even larger $M_X$, the production rate of $X$ would decrease, decreasing the abundance of $X$ (and therewith the value of $\eta$) again.
In these examples, the largest number of $\nu_R$ is produced in the final $X$ decay, which also generates these $\nu_R$ with a large momentum $p\sim M_X$ relative to the cooled-down SM bath, which leads to fairly large $\Delta N_\text{eff}$ in both examples.
Decreasing $\Gamma_X$, i.e.~increasing the $X$ lifetime, would not change $\eta$ significantly, but $\Delta N_\text{eff}$ would increase proportional to $1/\sqrt{\Gamma_X}$ due to the increased $\nu_R$ momentum relative to the SM bath temperature, following Eq.~\eqref{eq:superWIMP_Neff}.
Changing $B_R$ also does not affect $\eta$ in this region of parameter space, although it significantly affects $\Delta N_\text{eff}\propto B_R$.

Case (I), i.e.~the parameter space with $\Gamma_X\ll \mathcal{H}(M_X)$ is relatively easy to describe since $\eta$ only depends on $M_X$ and $\Delta N_\text{eff}$ follows from Eq.~\eqref{eq:superWIMP_Neff}.
Once we increase $\Gamma_X$ to values around $\mathcal{H}(M_X)$, the Boltzmann equations become more difficult due to several competing rates.

\begin{figure*}
    \centering
    \includegraphics[scale=0.7]{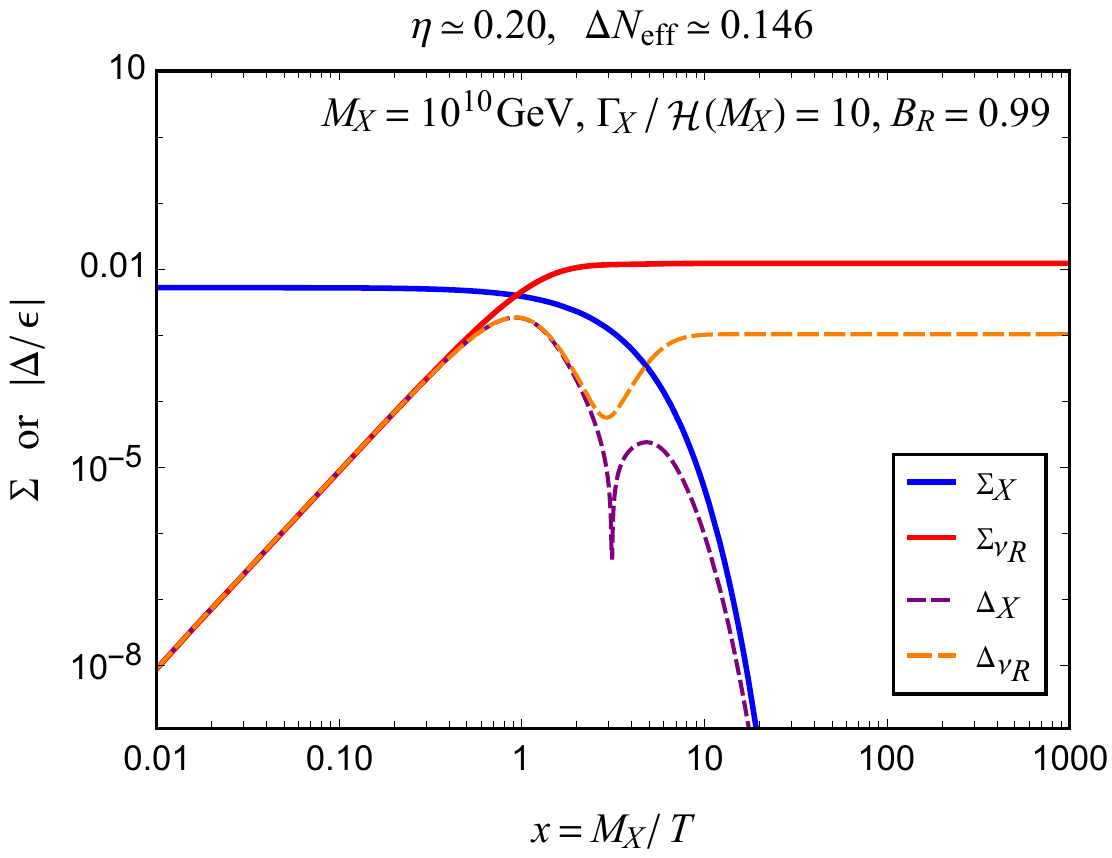} \hspace{10mm}
    \includegraphics[scale=0.7]{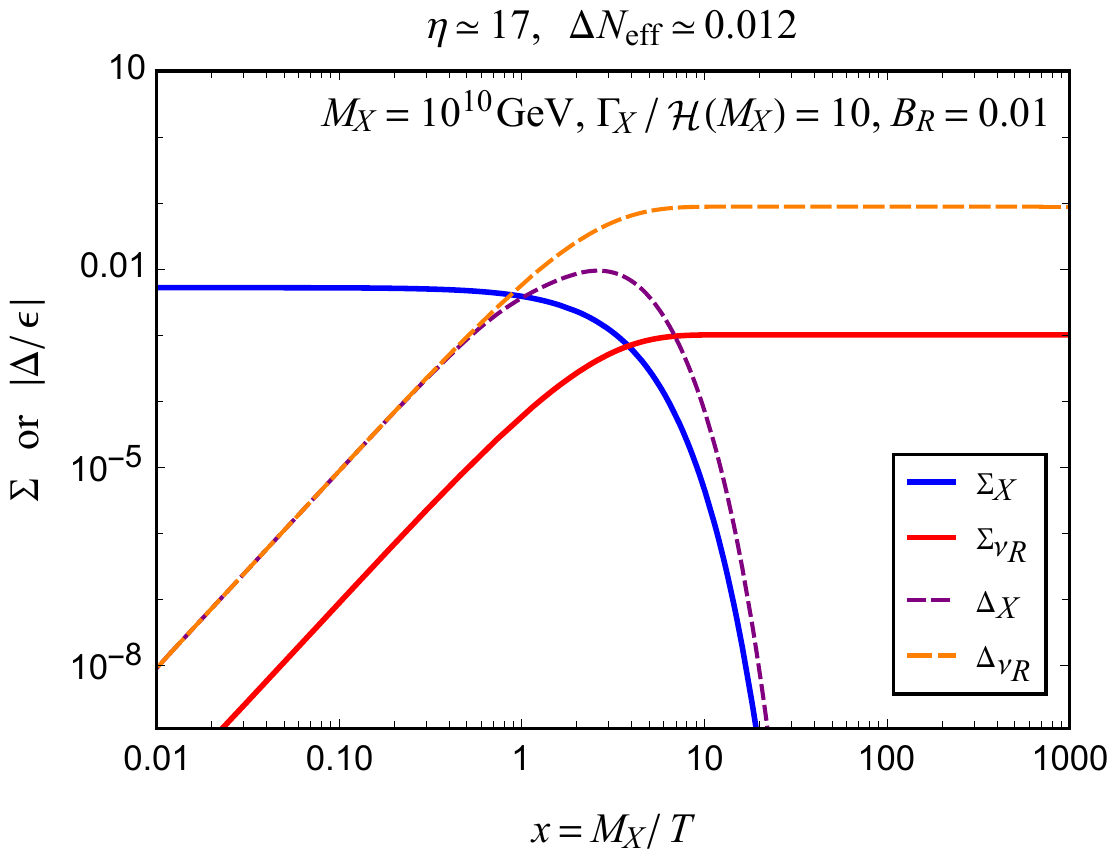}\\
    \caption{Evolution of $\Sigma_{X,\nu_R}$ and $|\Delta_{X,\nu_R}/\epsilon|$ for two parameter points together with the resulting $\eta$ and $\Delta N_\text{eff}$. Note that $\Sigma_X$ is virtually in equilibrium in the considered $x$ range.
    }
    \label{fig:evolution2}
\end{figure*}

The two plots in Fig.~\ref{fig:evolution2} are for $\Gamma_X/\mathcal{H}(M_X) = 10$ with $B_R=0.99$ (left) and $B_L=0.99$ (right). In the left plot, the $X\leftrightarrow e_R \nu_R$ rates are in equilibrium but the $X\leftrightarrow e_L\nu_L$ rates are not [case (II)], whereas the roles are reversed in the right plot [case (III)].
Similar to triplet leptogenesis, one rate being out of equilibrium is sufficient for a large $\eta$ despite $X$ being virtually in equilibrium. In the left plot, the $\nu_R$ reach equilibrium and give a thermal $\Delta N_\text{eff}$ [Eq.~\eqref{eq:thermal_Neff}], whereas the smaller $B_R$ in the right plot suppresses  $\Delta N_\text{eff}$.

\begin{figure*}[tb]
    \centering
    \includegraphics[scale=0.7]{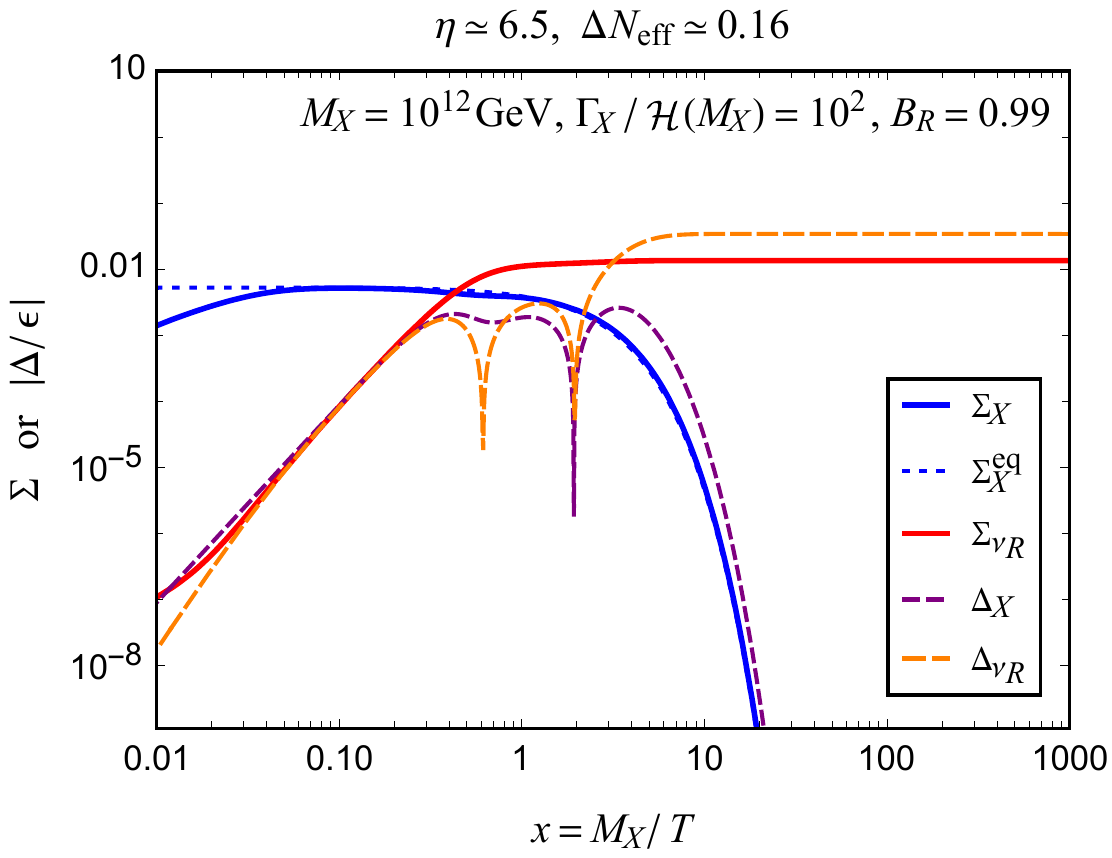} \hspace{10mm}
    \includegraphics[scale=0.7]{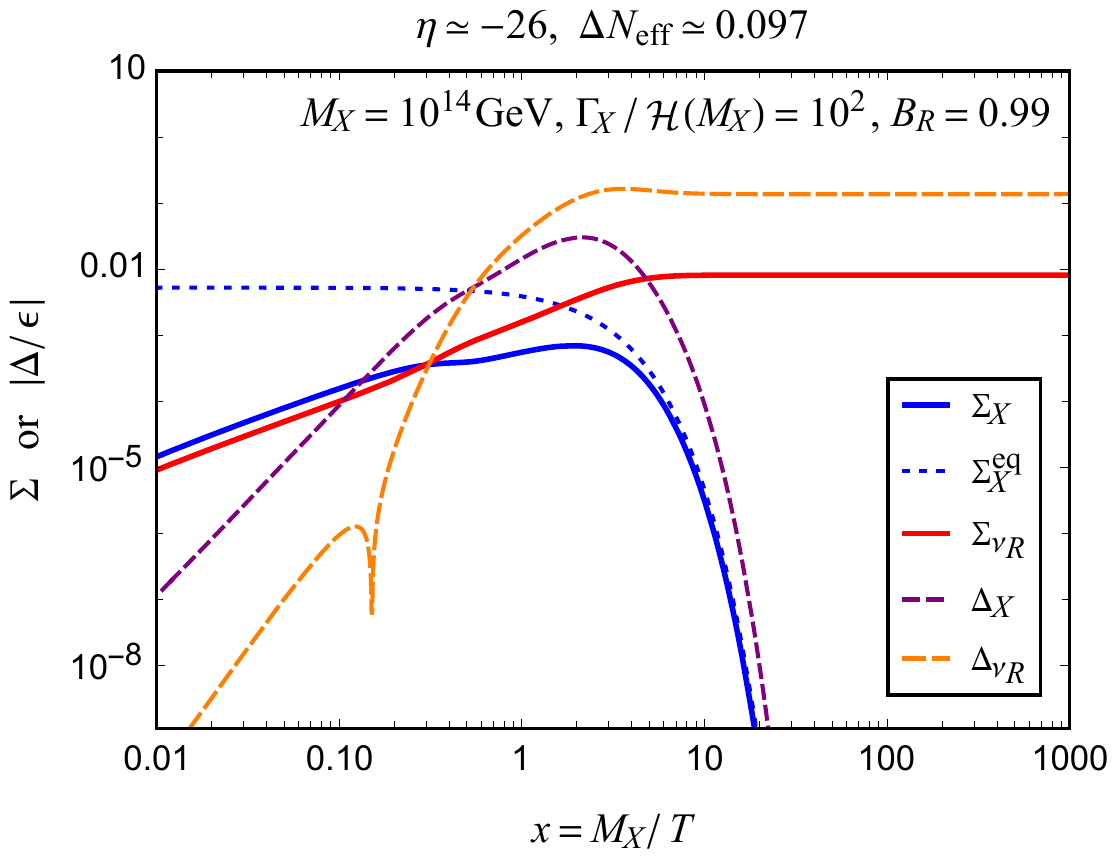}\\
    \caption{Evolution of $\Sigma_{X,\nu_R}$ and $|\Delta_{X,\nu_R}/\epsilon|$ for two parameter points together with the resulting $\eta$ and $\Delta N_\text{eff}$. The dotted blue line shows the equilibrium distribution of $\Sigma_X$.
    }
    \label{fig:evolution3}
\end{figure*}

Fig.~\ref{fig:evolution3} (left) is another illustration of case (II) with even larger $\Gamma_X$. Here, the gauge interactions and $\nu_R$ decay rates are strong, but the $X\leftrightarrow e_L\nu_L$ rates are just on the verge of equilibrium: $B_L \Gamma_X/\mathcal{H}(M_X)=1$. 
This is still sufficient for a very effective asymmetry generation. Increasing $\Gamma_X$ further would lead to a decreasing $\eta$ since the $X\leftrightarrow e_L\nu_L$ would thermalize and wash out the asymmetry.
Notice that the large $\nu_R$ rates lead to a  $\Delta N_\text{eff}$ that is slightly larger than the thermal value. The difference is small though, much larger values for  $\Delta N_\text{eff}$ can only be obtained for $\Gamma_X/\mathcal{H}\ll 1$.

Fig.~\ref{fig:evolution3} (right) is again case (II) but has a large $M_X$ and thus a much smaller $X$ annihilation rate, leading to out-of-equilibrium $X$. This leads to an even larger $|\eta|$, notably with a different sign than in Fig.~\ref{fig:evolution3} (left).

\clearpage

\bibliographystyle{utcaps_mod}
\bibliography{bib.bib}

\end{document}